\documentclass[prd,aps,twocolumn,nofootinbib,showpacs,amsmath,superscriptaddress]{revtex4-1}
\usepackage{graphicx}
\usepackage{epstopdf}
\usepackage{bm}
\usepackage{epsfig}
\usepackage{graphics}
\usepackage{xspace}
\usepackage{amssymb}
\usepackage{amsmath}
\usepackage{xcolor}
\usepackage[utf8]{inputenc}
\usepackage[normalem]{ulem}
\usepackage{stmaryrd}

\usepackage{comment}

\def\nslash{n\!\!\!\slash}

\def\OMIT#1{}

\newcommand{\nn}{\nonumber}

\newcommand{\bn}{{\bar n}}
\newcommand{\bea}{\begin{eqnarray}}
\newcommand{\eea}{\end{eqnarray}}

\newcommand{\gsim}{\mathrel{\rlap{\lower4pt\hbox{\hskip1pt$\sim$}}\raise1pt\hbox{$>$}}}

\newcommand{\be}{\begin{equation}}
\newcommand{\ee}{\end{equation}}

\newcommand{\red}[1]{\textcolor{black}{#1}}

\begin{document}



\title{\bf  Jet Charge with Global Event Shapes: Probing Quark Flavor Dynamics}

\author{Yang-Ting Chien}
\affiliation{
Physics and Astronomy Department, Georgia State University, Atlanta, GA 30303
}
\author{ Sonny Mantry}
\affiliation{Department of Physics and Astronomy, 
                   University of North Georgia,
                  Dahlonega, GA 30597, USA}



\newpage


\begin{abstract}
  \vspace*{0.3cm}

We propose measuring the jet electric charge of jet regions, defined within the framework of global event shapes, as a probe of quark flavor dynamics within the nucleon and the hadronization process. 
In particular, we consider a measurement of the jet region charge while simultaneously keeping track of the energy flow throughout the event, as characterized by the global event shape. 
As a concrete example, we focus on the measurement of the \textit{1-Jettiness jet charge} (${\cal Q}$),  the jet charge of the jet region ($J$) defined within the  framework of the 1-Jettiness global event shape ($\tau_1$) for the Deep Inelastic Scattering (DIS) process, $e^- + p \to e^- + J + X$, with unpolarized or longitudinally polarized protons. 
The 1-Jettiness distribution, binned according to jet charge, 
allows for  enhanced quark flavor separation of the initial state unpolarized or polarized PDFs. On the other hand, the jet charge distribution binned by 1-Jettiness can serve as a probe of quark flavor dynamics in the final state hadronization process. We derive a factorization theorem for simultaneous measurements of  $\tau_1$ and ${\cal Q}$ in the resummation region, $\tau_1 \ll P_{J_T}$, where  $P_{J_T}$ denotes the transverse momentum of the jet region.
The correlation between the jet region charge and the charge of the struck quark is expected to be the strongest in the resummation region. The factorization theorem contains a new universal 
\textit{charged jet function}, generalizing the standard jet function to include a jet charge measurement. Therefore these universal functions can be extracted from a global analysis of  N-jettiness and thrust at $e^+e^-$ colliders. We provide simulation studies to demonstrate the sensitivity of the 1-Jettiness jet charge observable to quark flavor dynamics in nucleon structure and explore the possibility of probing the final state hadronization process. This observable is well-suited for applications with existing HERA data and the future Electron-Ion Collider (EIC).

\end{abstract}

\maketitle

\newpage

\section{Introduction}

Global Event shapes are infrared and collinear safe observables that characterize the geometric pattern of hadronic energy flows in the final state. They act as a continuous measure of how pencil-like, planar, or spherical the ``shape" of the final state radiation is. 
Examples of global event shapes include thrust~\cite{Farhi:1977sg,Dasgupta:2002dc}, heavy jet mass~\cite{Clavelli:1979md}, 
C-parameter~\cite{Parisi:1978eg}, Broadening~\cite{Georgi:1977sf},  N-Jettiness~\cite{Stewart:2010tn}, and 1-Jettiness~\cite{Kang:2012zr,Kang:2013wca,Kang:2013nha,Kang:2013lga,Kang:2014qba,Chu:2022jgs,Cao:2024ota,Ee:2025scz} for Deep Inelastic Scattering (DIS). They have been studied in a variety of applications, including precision extractions of the strong coupling constant $\alpha_s$\red{~\cite{Becher:2008cf, Abbate:2010xh,Abbate:2012jh,Bell:2023dqs,Benitez:2024nav,Benitez:2025vsp}} and the top quark mass\red{~\cite{Fleming:2007qr,Fleming:2007xt,Hoang:2017kmk,Bachu:2020nqn,Dehnadi:2023msm}}. Such global event shapes have also been computed to very high accuracy and compared to data \cite{Becher:2008cf,Chien:2010kc,Hoang:2025uaa}, providing important tests of perturbative QCD and hadronization models. \red{Recently, there have also been many developments in the use of energy-energy correlators (EEC)~\cite{Basham:1979gh,Basham:1978zq,Basham:1978bw,Basham:1977iq} as a probe of QCD and hadronization, complementing traditional global event shapes. Detailed and comprehensive reviews of EECs can be found in Refs.~\cite{Neill:2022lqx,Moult:2025nhu}.}

Due to the light quark flavor symmetry of QCD, global event shapes do not provide much information on light quark flavor dynamics in initial state nucleon structure or final state hadronization. Typically, a global QCD analysis over a wide range of observables is employed to extract universal initial state structure functions and final state fragmentation functions, separated by flavor. These include inclusive observables as well as more exclusive ones that involve tagging a final state hadron. For example, the neutral current and charged current fully inclusive deep inelastic scattering (DIS) processes, $e^-+p\to e^-+X$ and $e^-+p\to \nu_e+X$, respectively,  are sensitive to different flavor combinations of initial state parton distribution functions (PDFs). Similarly, using polarized protons in these processes allows for disentangling initial state helicity PDFs by flavor. More exclusive observables, such as semi-inclusive DIS (SIDIS), $e^-+p\to e^-+ h+X$, where a final state hadron, $h$, is tagged, allows for additional flavor separation of initial state nucleon structure functions. Additionally, it acts as a probe of fragmentation functions by flavor. \red{Tagging final state hadrons also allows for measuring charge-charge correlations~\cite{Monni:2025zyv} or EECs of charged particles~\cite{Lee:2023npz}.} Measurements at the Relativistic Heavy Ion Collider (RHIC) and Large Hadron Collider (LHC) have shown that jets can also be a powerful probe of nucleon structure, including nucleon spin structure~\cite{Aschenauer:2016our,Adamczyk:2017wld,Boer:2014lka}.  Such jet based probes have now also been developed for use at the  EIC ~\cite{Liu:2018trl,Arratia:2020nxw,Liu:2020dct,Aschenauer:2019uex,Arratia:2019vju,Arratia:2020ssx,Kang:2020xyq,Aschenauer:2017jsk,Zheng:2018ssm}. However, without tagging final state hadrons, flavor separation remains challenging in inclusive jet measurements.  

The idea of jet charge was first explored~\cite{Field:1977fa} more than four decades ago. More recently, the jet electric charge has been shown~\cite{Krohn:2012fg,Waalewijn:2012sv} to be able to provide discrimination between jets initiated by up-type and down-type quarks at the LHC. Jets are initiated by energetic partons emerging from a hard collision process, subsequently producing a parton shower through perturbative soft and collinear emissions which is followed by hadronization into a cluster of hadrons. A suitable definition of the measured jet charge will correlate with the electric charge of the jet-initiating parton, providing a jet tagging tool that discriminates between jets initiated by different parton flavors. Jet charge has been used for DIS studies~\cite{Berge:1979qg,Berge:1980dx,Allen:1982ze,Albanese:1984nv,Barlag:1981wu,Erickson:1979wa},  measurements of the charge asymmetry~\cite{Stuart:1989db,Decamp:1991se},   tagging $b$-quarks in jets~\cite{Braunschweig:1990cv,Abazov:2006vd,Aaltonen:2013sgl,Aad:2013uza} and hadronically decaying $W$ bosons~\cite{Khachatryan:2014vla,Chen:2019uar},  probing anomalous Z~\cite{Li:2021uww} and Higgs~\cite{Wang:2023azz} boson couplings, discriminating between Higgs production mechanisms~\cite{Li:2023tcr},  testing aspects of perturbative~\cite{Krohn:2012fg,Waalewijn:2012sv, Aad:2015cua,Sirunyan:2017tyr,Fraser:2018ieu} and nonperturbative~\cite{Kang:2023ptt} Quantum Chromodynamics (QCD), probing nuclear-medium-induced jet quenching effects in heavy ion collisions~\cite{Gyulassy:1993hr,Wang:1994fx,Chien:2016led,Connors:2017ptx,Li:2019dre,Sirunyan:2020qvi,Du:2020pmp}, and probing the quark flavor structure of unpolarized and polarized~\cite{Kang:2020fka,Kang:2020xez} transverse momentum dependent parton distribution functions (TMDPDFs). The jet charge has also been used as one of a variety of  jet substructure tools~\cite{Gallicchio:2011xq,Larkoski:2017jix,Sirunyan:2020qvi,Brewer:2020och} for jet discrimination~\cite{Chien:2018dfn,Buckley:2020kdp,Kang:2021ryr}, with enhanced performances using machine learning techniques~\cite{Fraser:2018ieu,Chen:2019uar}. 

The standard definition of the jet charge~\cite{Krohn:2012fg,Waalewijn:2012sv} for a jet with transverse momentum, $P_{J_T}$, is given by:
\bea
\label{jc}
{\cal Q}_\kappa = \sum_{h\in {\rm jet}}  z_h^\kappa \> Q_h,
\eea
where the sum is over all the hadrons in the jet, $Q_h$ denotes the electric charge of the hadron, $h$, and \red{$z_h=p_{T_h}/P_{J_T}$ or $z_h=E_{h}/E_{J}$ for $pp$ and $ep$ or $e^+e^-$ colliders, respectively. These correspond to  the  jet transverse momentum fraction relative to the beam axis or jet energy fraction carried by the hadron.} The parameter $\kappa >0$ ensures that the contribution to the jet charge from soft hadrons is suppressed, ensuring infrared safety. 
It was shown that $\kappa\sim0.3$ provides the best discrimination~\cite{ATLAS:2013cf} between jets initiated by different quark flavors and will be the default value used in our analysis. A modified definition of the standard jet charge, the Dynamic Jet Charge~\cite{Kang:2021ryr}, allows for the $\kappa$ parameter to be a function of some property of the jet constituents. It can be chosen as a function of $z_h$, $\kappa=\kappa(z_h)$, to improve discrimination between quark and gluon jets and  make jet discrimination more resilient against the contamination from underlying events. We will often drop the $\kappa$ subscript in Eq.~(\ref{jc}) and refer to the jet charge as simply, ${\cal Q}$, for brevity in notation. In our analysis, we consider both the standard and dynamic jet charge definitions, as they have different jet discrimination capabilities and sensitivity to soft radiation, serving as complementary probes.
 
Recently, it was shown that the jet charge property can be used for flavor separation of the unpolarized and polarized Transverse Momentum Dependent PDFs (TMDPDFs)~\cite{Kang:2020fka}. The unpolarized TMDPDFs describe the 3D momentum space distribution of partons in the nucleon. The polarized TMDPDFs, in addition, encode quantum correlations between the parton momentum distributions and the proton spin. The electron-proton scattering process, $e^-+p\to e^-+J+X$, in the limit of small electron-jet transverse momentum imbalance, $q_T=|\vec{p}_T^{\>e}+\vec{P}_{J_T}|\ll\vec{p}_T^{\>e} \sim P_{J_T}$, \red{defined relative to the beam axis in the center of mass frame,} is sensitive to TMDPDFs. The factorization formula for the $q_T$-distribution is sensitive to a particular flavor combination of TMDPDFs, a universal soft function, and a universal jet function. The jet function describes the dynamics of the leading jet that recoils against the electron, and in general depends on the jet radius, $R$, and the jet clustering algorithm. Flavor separation of the TMDPDFs is achieved through a jet charge measurement on the leading jet and binning the $q_T$-distribution by jet charge. The positive and negative jet charge bins provide enhanced sensitivity to the $u$-quark and the $d,s$ quark TMDPDFs, respectively, compared to the $q_T$-distribution integrated over all jet charge values. 

It was also recently explored that, charge and flavor correlation among hadrons is sensitive to details of underlying hadronization processes \cite{Chien:2021yol} therefore it can be used to test hadronization models in Monte Carlo implementations. While the studies require tagging individual hadron flavors, i.e. particle identifications, jet charge or subjet charges could provide more inclusive information about quark flavors. 

In this work, we build on these ideas and propose a new class of observables that incorporates the flavor discrimination power of jet charges within the framework of global event shapes as a probe of quark flavor dynamics in nuclear structure and hadronization. We propose measuring the jet charge (${\cal Q}$) of jet regions as defined by the global event shape. 
This allows for measuring the jet region charge while simultaneously keeping track of the energy flow throughout the event, as quantified by the global event shape. In particular, one can bin event shape distributions according to the jet region charge. Conversely, one can bin jet region charge distributions according to the global event shape value. Such observables can allow for disentangling flavor dynamics in the initial state nucleon structure functions and in the final state hadronization process. 

We focus on the 1-Jettiness~\cite{Kang:2012zr,Kang:2013wca,Kang:2013nha,Kang:2013lga,Kang:2014qba,Chu:2022jgs,Cao:2024ota} global event shape for the neutral current DIS process,  $e^- + p \to e^- + J + X$.  The transverse momentum and rapidity of the jet region, $J$, are  denoted by $P_{J_T}$ and $y_J$, respectively. The usual 1-Jettiness cross section, differential in $(\tau_1, P_{J_T},y_J)$, is denoted by:
\bea
\label{eq:obs0}
d\sigma \left [\tau_1, P_{J_T}, y_J \right ] \equiv \frac{d^3\sigma (e^- + p \to  J + X)}{dy_J\> dP_{J_T}\>d\tau_1}.
\eea

The resummation region of the $\tau_1$-distribution, $\tau_1 \ll P_{J_T}$, corresponds to events with a single jet topology such that energetic radiation, $E\sim P_{J_T}\gg \tau_1$, is collimated either along the proton beam or the leading jet direction, and only soft radiation, $E\sim \tau_1 \ll P_{J_T}$ at wide angles from the beam and jet directions is allowed. It requires resummation of large Sudakov logarithms of the form $\alpha_s^n \ln ^{2m} (\tau_1/P_{J_T})$, for $m\leq n$, arising from the $\tau_1\ll P_{J_T}$ restriction on final state radiation. This restriction effectively acts as a veto on additional jets at wide angles from the leading jet and beam directions. In this resummation region, the differential cross section has the schematic, factorized form
\bea
\label{schem-1}
d\sigma_{\rm resum} \left [\tau_1, P_{J_T}, y_J \right ] \sim H\otimes B \otimes J \otimes {\cal S},
\eea
where $H$ denotes the hard function describing the hard partonic scattering, $J$ denotes the jet function describing the energetic radiation collinear with jet region, ${\cal S}$ denotes the soft function describing the soft radiation throughout the event, and $B$ denotes the beam function~\cite{Stewart:2009yx} describing initial state collinear radiation along the beam. The beam function can be further factorized as~\cite{Stewart:2010tn,Stewart:2009yx}
\bea
\label{schem-2}
B\sim {\cal I}\otimes f,
\eea
where ${\cal I}$ describes the perturbatively calculable initial state collinear radiation and $f$ denotes the PDF. 

On the other hand, the fixed-order region,  $\tau_1 \sim P_{J_T}$, corresponds to events with additional energetic radiation at wide angles from the beam and leading jet directions. 
In this region, calculations are done in the fixed-order perturbative QCD framework. The resummation and fixed-order regions can be matched to give a smooth 1-Jettiness spectrum over the entire range of $\tau_1$. Most recently, results for the 1-Jettiness spectrum was obtained at the N$^3$LL+${\cal O}(\alpha_s^2)$ level of accuracy~\cite{Cao:2024ota}.

We extend this 1-Jettiness framework by, in addition, measuring the charge, ${\cal Q}$, of the jet region, $J$. We refer to it as the ``\textit{1-Jettiness Jet Charge}." The cross section is now differential in four variables, $({\cal Q}, \tau_1,  P_{J_T},y_J)$, and is denoted as:
\bea
\label{eq:obs0QJ}
d\sigma \left [{\cal Q},\tau_1, P_{J_T}, y_J \right ] \equiv \frac{d^3\sigma (e^- + p \to  J + X)}{dy_J\> dP_{J_T}\>d\tau_1\>d{\cal Q}}.
\eea
The correlation between the 1-Jettiness jet charge, ${\cal Q}$, with the electric charge of the hard-scattered quark, allows for enhanced flavor separation of the initial state PDFs through the jet-charge-binned 1-Jettiness distributions. This correlation is expected to be the strongest in the resummation region, $\tau_1\ll P_{J_T}$, where the struck quark evolves into the final state jet region through collinear and soft splittings without any hard emissions at wide angles. In this resummation region, using the Soft-Collinear Effective Theory (SCET)~\cite{Bauer:2000ew,Bauer:2000yr,Bauer:2001ct,Bauer:2001yt,Bauer:2002nz,Beneke:2002ph}, the differential cross section for the 1-Jettiness Jet Charge distribution has the schematic form
\bea
\label{schem-3}
d\sigma_{\rm resum} \left [{\cal Q}, \tau_1, P_{J_T}, y_J \right ] \sim H\otimes B \otimes {\cal G} \otimes {\cal S},
\eea
which has the same form as Eq.~(\ref{schem-1}), except for the replacement $J\to {\cal G}$, corresponding to replacing the standard jet function with a \textit{jet charge jet function}. This formula will receive power suppressed contributions from the contribution of soft radiation to the jet region charge, as discussed in appendix~\ref{PC}. The jet charge jet function, ${\cal G}$, is obtained through the insertion of a jet charge measurement operator in the standard jet function. The new jet charge jet function ${\cal G}$ is in general nonperturbative since the jet charge measurement is made on final state hadrons, making it sensitive to the hadronization process. However, it is universal therefore it can be extracted from other jet-based global event shapes such as the N-Jettiness~\cite{Stewart:2010tn} global event shape at $e^+e^-$ colliders or the LHC. 
The differential cross section in Eq.~(\ref{eq:obs0QJ}) can be integrated over ${\cal Q}$-bins to give the jet-charge-binned 1-Jettiness distribution. In the resummation region, the jet-charge-binned 1-Jettiness distribution  has the schematic form
\bea
\label{schem-3-Bin}
d\sigma_{\rm resum} \left [{\cal Q}^{\rm bin}, \tau_1, P_{J_T}, y_J \right ] \sim H\otimes B \otimes {\cal G}_{\rm bin} \otimes {\cal S},
\eea
where ${\cal G}_{\rm bin}$ denotes the jet charge function integrated over a jet charge bin. The positive jet charge bin (${\cal Q}_J>0.25$) will give more weight to the contribution of the $u$-quark to the 1-Jettiness spectrum, corresponding to its positive electric charge. The negative jet charge bin (${\cal Q}_J<-0.25$) will give more weight to the contribution of the $d,s$-quarks to the 1-Jettiness spectrum, corresponding to their negative electric charge. Thus, binning the 1-Jettiness distribution according to the 1-Jettiness jet charge allows for disentangling initial state proton structure functions by flavor. For scattering of unpolarized protons, this corresponds to separating the collinear PDFs by flavor. The longitudinal single spin asymmetry constructed from scattering off polarized protons will allow for flavor separation of helicity PDFs.

Alternatively, with the multi-differential cross section in Eq.~(\ref{eq:obs0QJ}) one can integrate over $\tau_1$-bins to give the 1-Jettiness-binned jet charge distribution. This allows one to study the jet charge distribution as a function of the shape of energy flow in the final state as characterized by the value of $\tau_1$. Such studies can probe flavor dynamics in the hadronization process, providing a new tool for distinguishing between different hadronization models.
 
We also note that this observable can be adapted to the charged current DIS events, $e^- + p \to \nu_e + J + X$ since the undetected neutrino, tagged as missing energy, does not contribute to the $\tau_1, {\cal Q}, P_{J_T},$ or $y_J$ observables.  Such a charge current DIS observable will complement the analysis of neutral current DIS as it probes different combinations of initial state proton structure functions, allowing for further separation of structure functions by flavor. We leave the charge current DIS analysis for future work. 
 
Finally, we note that the formalism for the 1-Jettiness charge observable developed here is ideally suited for the proposed Electron-Ion Collider (EIC) and can also be readily applied to the existing HERA data.
 
\section{1-Jettiness Global Event Shape}

For completeness,  we give a brief review of the formalism and notation for the 1-Jettiness global event shape for DIS which has been extensively studied~\cite{Kang:2012zr,Kang:2013wca,Kang:2013nha,Kang:2013lga,Kang:2014qba,Chu:2022jgs,Cao:2024ota}. For more detailed explanations of the relevant notation used in this work, see Ref.~\cite{Cao:2024ota}. We consider the neutral current (NC) DIS process:
\bea
e^{-}(k) + p(P) \to e^{-}(k') + J +X,
\eea
in the center of mass frame, where the initial state proton and electron momenta, $P^\mu$ and $k^\mu$ respectively,  are given by
\bea
\label{eq:nB}
P^\mu &=& \frac{\sqrt{s}}{2} n_{\red{B}}^\mu, \qquad n_{\red{B}}^{\mu}=(1,\red{\vec{n}_B}),\nn \\
k^\mu &=& \frac{\sqrt{s}}{2} \bn_{\red{B}}^\mu, \qquad \bn_{\red{B}}^{\mu}=(1,-\red{\vec{n}_B}),
\eea
\red{where $\vec{n}_B=\hat{z}$ is a unit vector point along the $z$-axis in the proton beam direction.}
The 1-Jettiness global event shape, $\tau_1$, is defined as
\bea
\label{tau1}
\tau_1 = \sum_{k} \min \Big \{ \frac{2q_B\cdot p_k}{Q_B}, \frac{2q_J\cdot p_k}{Q_J}\Big \},
\eea
where the sum is over all final state particles except the scattered electron. The $q_B^\mu$ and $q_J^\mu$ denote the beam and jet reference vectors, respectively, whose explicit expressions will be given shortly below. The $Q_B$ and $Q_J$ are energy scales at the order of the hard scale. Different choices for these quantities correspond to different definitions of 1-Jettiness. Each final state particle of momentum $p_k$ is grouped with either the beam region or the jet region according to the minimization condition in Eq.~(\ref{tau1}), and accordingly contributes to $\tau_1$. \red{In particular, the jet region is defined as the set of particles with momenta, $p_k$, that satisfy $2q_J\cdot p_k/Q_J< 2q_B\cdot p_k/Q_B$. Correspondingly,  the total jet momentum of the jet region is given by
\bea
\label{eq:PJtau1}
P_J^\mu = \sum_k p_k^\mu \> \theta \left (\frac{2q_B\cdot p_k}{Q_B} -\frac{2q_J\cdot p_k}{Q_J} \right ).
\eea} 
The largest contributions to $\tau_1$ come from final state particles with large energies, and at wide angles from both the beam and jet reference vectors. On the other hand, energetic particles closely aligned with either the beam or jet reference vectors give relatively small contributions to the $\tau_1$ value. Thus, events with small $\tau_1$ values have energetic radiation closely aligned with the beam or jet reference vectors. Events with large $\tau_1$ have additional energetic radiation at wide angles from both the beam and jet reference vectors. Therefore, $\tau_1$ quantifies the pattern of final state energy flow.

\begin{figure}
\centering
\includegraphics[scale=0.15]{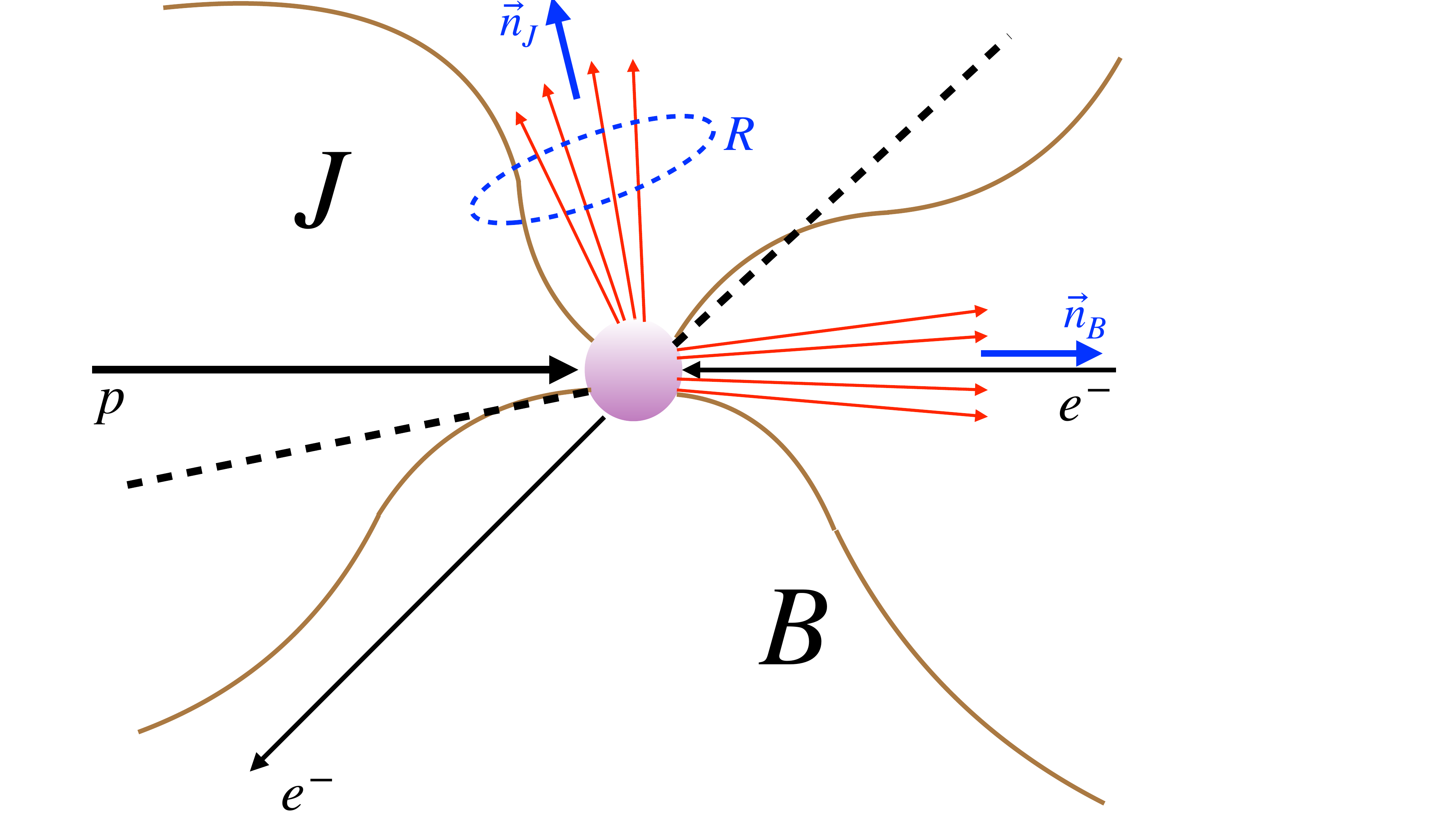}
    \caption{\red{Schematic illustration of the jet ($J$) and beam ($B$) regions, separated by the dashed lines, as determined by the 1-Jettiness algorithm. The final state energetic (red) and soft (brown) hadronic particles  are clustered into either the jet or the beam region, according to the 1-Jettiness algorithm. The jet radius, $R$, refers to the jet algorithm used to determine the jet reference vector, $q_J^\mu$.}
}
\label{fig:jet-region}
\end{figure}

\red{We choose the beam reference vector}
\bea
q_B^\mu&=& x P^\mu \red{= \omega_B \frac{n_B^\mu}{2}},
\eea
aligned with the beam direction. \red{Here $x$ denotes the proton momentum fraction carried by the parton participating in the hard interaction, and from Eq.~(\ref{eq:nB}) we have $\omega_B =x\sqrt{s}$.} The jet reference vector is chosen as
\bea
q_J=(K_{J_T} \cosh y_K, &\vec{K}_{J_T}&, K_{J_T} \sinh y_K), 
\eea
where $K_J^\mu$ is the four-momentum of the leading jet \red{defined by a standard jet algorithm. In our analysis, we have used the anti-k$_t$ algorithm with a jet radius of $R=1$.} Here $K_{J_T}$ and $y_K$ denote the transverse momentum and rapidity of the leading jet, and are used to construct the null jet reference vector, $q_J^\mu$. \red{Since $q_J^\mu$ is a null vector, it can be written as
\bea
\label{eq:nJ}
q_J^\mu =\omega_J\frac{n_J^\mu}{2}, \qquad n_J^\mu = (1,\vec{n}_J),
\eea
where $\vec{n}_J$ is a unit vector along the leading jet direction, and $\omega_J=2K_{J_T} \cosh y_K$.
}

\red{Different choices of jet algorithms only affect the determination of the reference vector $q_J^\mu$. The impact of  different choices is power suppressed in the resummation region, $\tau_1\ll P_{J_T}$, corresponding to events with the topology of a single jet, but can lead to order one differences in the fixed-order region $\tau_1\sim P_{J_T}$, characterized by more than one jet. The jet algorithm with jet radius $R$, is only used to determine the jet reference vector, $q_J^\mu=\omega_J n_J^\mu/2$. The final set of particles in the jet region is then determined by using this $q_J^\mu$ in Eq.~(\ref{eq:PJtau1}). }  

We choose the constants $Q_B$ and $Q_J$ as
\bea
Q_B&=&\omega_B, \qquad Q_J= \omega_J.
\eea
\red{For this choice of $Q_B$ and $Q_J$, the definition of $\tau_1$ reduces to the simpler form
\bea
\label{tau1nBnJ}
\tau_1 = \sum_{k} \min \Big \{ 2n_B\cdot p_k, \>2n_J\cdot p_k\Big \},
\eea
so that the choice of the beam and jet reference vectors corresponds to the choice of directions $\vec{n}_B$ and $\vec{n}_J$, since the time-like components are fixed to be $n_B^0=n_J^0=1$, as seen from Eqs.~(\ref{eq:nB}) and (\ref{eq:nJ}). In this case, these directions could be determined through a minimization procedure, requiring that they minimize the value of $\tau_1$ in Eq.~(\ref{tau1nBnJ}). One might also consider fixing the direction of $\vec{n}_B$ to be along the proton beam axis, varying only the direction of $\vec{n}_J$ during the minimization procedure. We note that through such minimization procedures, it is possible  to  define $\tau_1$, and the corresponding jet and beam regions,  without the use of any jet algorithm. However, for convenience, in our analysis we have made use of a jet algorithm to determine $\vec{n}_J$, as described earlier.}
\red{Ultimately, differences in the methods used to define the jet reference vector $q_J^\mu/Q_J$ will yield power suppressed effects in the resummation region, $\tau_1\ll P_{J_T}$, so that the structure of the corresponding factorization theorem  does not explicitly depend on them at leading power in $\tau_1/P_{J_T}$.}

\begin{figure}
    \centering
    \includegraphics[scale=0.5]{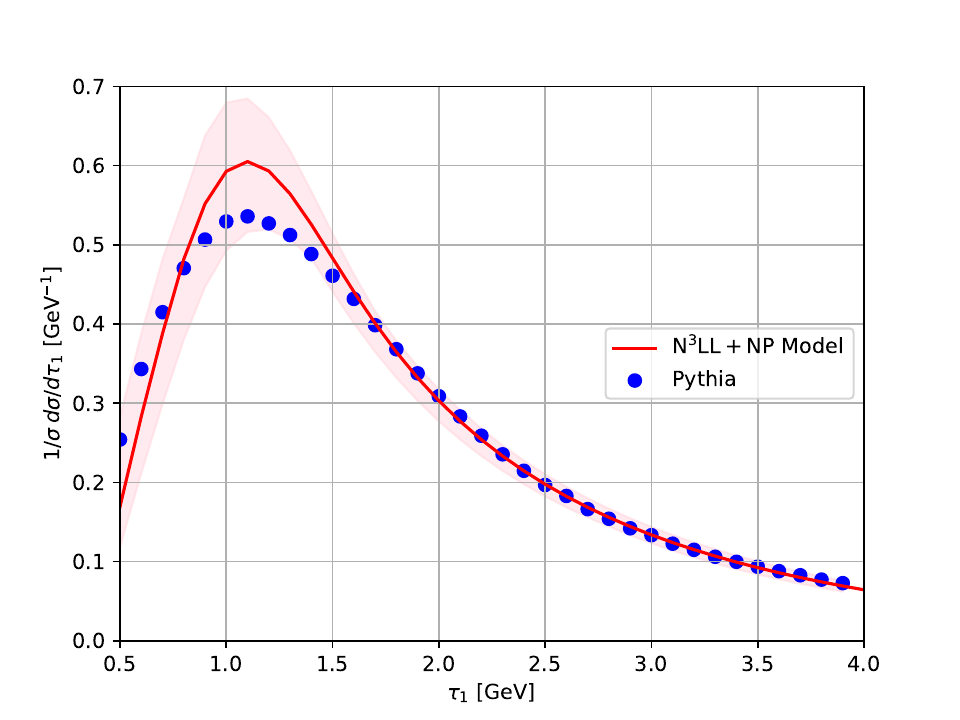}
    \caption{Pythia 8.312~\cite{Sj_strand_2015,Bierlich:2022pfr} simulation data (blue dots) for the normalized 1-Jettiness $\tau_1$-distribution compared to the theoretical prediction at the N$^3$LL level of accuracy with a soft function model that describes hadronization effects. The scale variations (pink band), describing the theoretical uncertainty, are normalized to the central curve (red) over the displayed range.  The relevant EIC kinematics chosen are: $\sqrt{s}=90.0$ GeV, $P_{J_T}= [20.0,30.0]$ GeV, and $y_J=[-2.5,2.5]$.
}
    \label{fig:1-jettiness-spec}
\end{figure}

The factorization formula for the $\tau_1$ distribution in the resummation region is given by~\cite{Cao:2024ota}
\bea
\label{eq:factorization_resum}
&& d\sigma_{\rm resum} \left [\tau_1,P_{J_T},y_J \right ] =\sigma_0 \>H(\xi, \mu; \mu_H) \nn \\
&&\times \int ds_J \int dt_a  \> {\cal S}\left(\tau_1 - \frac{t_a}{Q_a}-\frac{s_J}{Q_J}, \mu;\mu_S\right) 
\nn \\
&&\times \>\Big [ \sum_{q_i} L_{q_i}\> J^{q_i}(s_J, \mu;\mu_J)\> B_{q_i}(x_*,t_a,\mu;\mu_B)  \nn \\
&&+ \sum_{\bar{q}_i} L_{\bar{q}_i}\>J^{\bar{q}_i}(s_J, \mu;\mu_J)\> B_{\bar{q}_i}(x_*,t_a,\mu;\mu_B) \Big ] ,
\eea
\red{where $Q_{q_i}^2 \sigma_0$  is the partonic Born-level cross section due to single photon exchange for an electron scattering off a quark, $q_i$, with electric charge, $Q_{q_i}$, in units of the proton electric charge.  For single photon exchange, $L_{q_i}=Q_{q_i}^2$, but it can include additional terms corresponding to the relative  additional contributions arising from including $Z$-boson exchange. Explicit expressions for $\sigma_0$ and $L_{q_i,\bar{q}_i}$ can be found in Refs.~\cite{Cao:2024ota} and \cite{Kang:2013nha, Cao:2024ota}, respectively.}  $H$, ${\cal S}, J^{q_i,\bar{q}_i},$ and $B_{q_i,\bar{q_i}}$ denote the hard function, soft function, quark and antiquark jet functions, and quark and antiquark beam functiuons, respectively, and 
\bea
x_* = \frac{e^{y_J} P_{J_T}}{\sqrt{s}-e^{-y_J}P_{J_T}}, \>\>\>\> \xi =x_*  \sqrt{s} P_{J_T}e^{-y_J}. 
\eea
Note that the hard, soft, jet, and beam functions are all evaluated at the common renormalization scale $\mu$ after renormalization group evolution from the scales $\mu_H \sim P_{J_T}, \mu_S\sim\tau_1, \mu_J \sim \mu_B \sim \sqrt{P_{J_T}\tau_1}$ which eliminates large logarithms in each of the respective functions. The renormalization group evolution sums large Sudakov logarithms arising from ratios of these scales. 
The quark and anti-quark beam functions ($B_{q,\bar{q}}$) are matched~\cite{Stewart:2009yx}  onto the PDFs as
\bea
\label{beam}
B_{(q_i,\bar{q}_i)}(x,t_a,\mu;\mu_B) &=&\sum_i \int_x^1 \frac{dz}{z} {\cal I}_{({q_i,\bar{q}_i})j}  \left(\frac{x}{z}, t_a, \mu;\mu_B\right)  \nn \\
&&\times f_{j/A}(z,\mu_B), 
\eea
where the $ {\cal I}_{({q_i,\bar{q}_i})j}$ functions are perturbatively calculable.  The hard~\cite{Idilbi_2006,Becher:2006mr}, soft~\cite{Jouttenus:2011wh,Boughezal:2015eha}, jet~\cite{Becher:2006qw, Idilbi_2006,Becher:2006mr}, and beam functions~\cite{Stewart:2010qs,Gaunt:2014xga} are known to ${\cal O}(\alpha_s^2)$. \red{The jet function is now known~\cite{Bruser:2018rad} to ${\cal O}(\alpha_s^3)$.} Results for this 1-Jettiness spectrum are known~\cite{Cao:2024ota} at the N$^3$LL+${\cal O}(\alpha_s^2)$ level of accuracy, and shown in Fig.~\ref{fig:1-jettiness-spec} for typical EIC kinematics.

\section{1-Jettiness Jet Charge Factorization Framework}
The 1-Jettiness jet charge distribution corresponds to the measurement of the jet charge of the jet region as defined by the 1-Jettiness global event shape, in addition to the measurement of the 1-Jettiness global event shape itself. In the resummation region, $\tau_1\ll P_{J_T}$, its factorization formula is given by
\bea
\label{eq:factorization_resum_Q_J}
&&d\sigma_{\rm resum} \left [{\cal Q}, \tau_1,P_{J_T},y_J \right ] =\sigma_0 \> H(\red{\xi}, \mu; \mu_H) \nn \\
&&\times \int ds_J \int dt_a  \> {\cal S}\left(\tau_1 - \frac{t_a}{Q_a}-\frac{s_J}{Q_J}, \mu;\mu_S\right) 
\nn \\
&&\times \Big [ \sum_{q_i} \>L_{q_i}\>{\cal G}^{q_i}({\cal Q}, s_J, \mu;\mu_J)\>  B_{q_i}(x_*,t_a,\mu;\mu_B)  \nn \\
&&\>\>\>+ \sum_{\bar{q}_i} L_{\bar{q}_i}\>{\cal G}^{\bar{q}_i}({\cal Q}, s_J, \mu;\mu_J) \> B_{\bar{q}_i}(x_*,t_a,\mu;\mu_B) \Big ] .
\eea
Compared to Eq.~(\ref{eq:factorization_resum}), the jet function $J^q(s_J,\mu)$ is now replaced with the \textit{charged jet function}, ${\cal G}^{q}({\cal Q},s_J,\mu)$, which corresponds to the insertion of a jet charge measurement operator in the standard jet function. Its field-theoretic definition is given in Eq.~(\ref{Gqfield-theoretic}). Note that the jet charge contribution from soft particles in the jet region is power suppressed compared to that of the energetic collinear particles. This can be understood from the definition of jet charge in Eq.~(\ref{jc}) which weighs the charge of each particle by its jet energy fraction. The dynamic jet charge allows for the contribution of the soft particles to be further suppressed by choosing the $\kappa$ parameter as a dynamic function of the hadron energy fraction, $z_h$. Therefore, at leading power, the jet charge measurement operator only acts on the energetic collinear particles in the jet region. Consequently, only the jet function, $J^q(s_J,\mu)$, is affected and replaced with the jet charge jet function ${\cal G}^{q}({\cal Q},s_J,\mu)$. We discuss power suppressed corrections arising from soft particles in appendix~\ref{PC}.

The charged jet function, ${\cal G}^{q}({\cal Q},s_J,\mu)$, is in general a non-perturbative object since the jet charge measurement operator is not collinear safe and sensitive to the hadronization process. However, ${\cal G}^{q}({\cal Q},s_J,\mu)$ has the same renormalization scale dependence as $J^q(s_J,\mu)$. This follows from the renormalization scale independence of the cross section and the fact that 1-Jettiness Jet Charge cross section in Eq.~(\ref{eq:factorization_resum_Q_J}) can be obtained simply by replacing $J^q(s_J,\mu)$ with ${\cal G}^{q}({\cal Q},s_J,\mu)$ in Eq.~(\ref{eq:factorization_resum}). Integrating Eq.~(\ref{eq:factorization_resum_Q_J}) over all ${\cal Q}$ should give back the standard 1-Jettiness cross section in Eq.~(\ref{eq:factorization_resum}). In particular,  integrating ${\cal G}^{q}({\cal Q},s_J,\mu)$ over ${\cal Q}$ gives back the standard jet function
\bea
\label{Gqnorm}
\int d{\cal Q} \> {\cal G}^{q}({\cal Q},s_J,\mu) = J^q(s_J,\mu),
\eea
which also follows from Eqs.~(\ref{Jqfield-theoretic}) and (\ref{Gqfield-theoretic}). 
In position space, the jet function, $J^{q}(y,\mu)$, and charged jet function, ${\cal G}^{q}({\cal Q},y,\mu)$, are defined through the Fourier transforms
\bea
\label{FT}
J^{q}(s,\mu) &=& \int \frac{dy}{2\pi}\> e^{i y s}\>J^{q}(y,\mu), \nn \\
{\cal G}^{q}({\cal Q},s,\mu) &=& \int \frac{dy}{2\pi}\> e^{i y s}\>{\cal G}^{q}({\cal Q},y,\mu).
\eea
It is useful to define the ratio of the position space charged and standard jet functions
\bea
\label{RqRatio}
R^{q}({\cal Q},y) \equiv \frac{{\cal G}^{q}({\cal Q},y,\mu)}{J^{q}(y,\mu)}.
\eea
Therefore ${\cal G}^{q}({\cal Q},y,\mu)$ and  $J^{q}(y,\mu)$ are related through the multiplicative function $R^{q}(Q,y)$ as
\bea
\label{Rq}
{\cal G}^{q}({\cal Q},y,\mu) &=& R^{q}({\cal Q},y) J^{q}(y,\mu).
\eea
Note that $R^{q}({\cal Q},y)$ is renormalization scale independent due to the cancellation between the renormalization group running, which is multiplicative in position space, of ${\cal G}^{q}({\cal Q},y,\mu)$ and $J^{q}(y,\mu)$ in the ratio. The multiplicative relation in Eq.~(\ref{Rq}) becomes a convolution in momentum space. Using Eq.~(\ref{Rq}) in Eq.~(\ref{FT}), we get
\bea
\label{JCFconv}
{\cal G}^{q}({\cal Q},s,\mu) &=& \int ds' \> R^q({\cal Q},s-s')\> J^{q}(s',\mu), 
\eea
where $R^q({\cal Q},s)$ corresponds to the momentum space definition
\bea
\label{RqFT}
R^q({\cal Q},s) &=& \int \frac{dy}{2\pi}\> e^{iys}\>R^q({\cal Q},y) .
\eea
Combining Eqs.~(\ref{Gqnorm}) and (\ref{JCFconv}) leads to the normalization conditions
\bea
\label{Rqnorm}
\int d{\cal Q}\>R^q({\cal Q},s) = \delta(s), \>\> \int d{\cal Q}\>R^q({\cal Q},y) = 1.
\eea
From Eq.~(\ref{JCFconv}), we see that the function $R({\cal Q},s)$ acts as the  integral kernel that transforms the standard jet function, $J^q(s,\mu)$, distribution in the mass squared variable, `$s$', to the new distribution, ${\cal G}^{q}({\cal Q},s,\mu)$, where the final state is restricted to, in addition, have a jet charge value of ${\cal Q}$. It is a universal, non-perturbative function that can be extracted from data on other jet-based global event shapes, as explained in section ~\ref{univ}.

Using Eq.~(\ref{JCFconv}) in Eq.~(\ref{eq:factorization_resum_Q_J}), the resummed factorization formula differential in jet charge takes the form:
\bea
\label{eq:fac_resum_Q_J_Rq}
&&d\sigma_{\rm resum} \left [{\cal Q}, \tau_1,P_{J_T},y_J \right ] =\sigma_0 \> H(\xi, \mu; \mu_H) \nn \\
&&\times \int ds_J\>\int dt_a \> \int ds_J' \>{\cal S}\left(\tau_1 - \frac{t_a}{Q_a}-\frac{s_J}{Q_J}, \mu;\mu_S\right) 
 \\
&&\times \Big [ \sum_{q_i} \>L_{q_i}\>R^{q_i}({\cal Q}, s_J-s_J')J^{q_i}(s_J', \mu;\mu_J)\>  B_{q_i}(x_*,t_a,\mu;\mu_B)  \nn \\
&&+ \sum_{\bar{q}_i} L_{\bar{q}_i}\>R^{\bar{q}_i}({\cal Q}, s_J-s_J') J^{\bar{q}_i}(s_J', \mu;\mu_J)\> B_{\bar{q}_i}(x_*,t_a,\mu;\mu_B) \Big ] . \nn
\eea
From the above factorization formula, we can also obtain the 1-Jettiness distribution integrated over a jet charge bin. Correspondingly, we can define the charged jet function integrated over a jet charge bin in momentum and position space as
\bea
\label{JCFconvBin}
{\cal G}^{q}_{\rm bin}(s,\mu) &=& \int ds'\> R^q_{\rm bin}(s-s')\>J^{q}(s',\mu), \nn \\
{\cal G}^{q}_{\rm bin}(y,\mu) &=&  R^q_{\rm bin}(y)\>J^{q}(y,\mu), 
\eea
where the binned $R^q_{\rm bin}(s)$ and $R^q_{\rm bin}(y)$ functions are given by
\bea
\label{RqBin}
R^q_{\rm bin}(s) &=& \int_{\rm bin} d{\cal Q} \>R^q({\cal Q} , s), \nn \\
R^q_{\rm bin}(y) &=& \int_{\rm bin} d{\cal Q}  \>R^q({\cal Q} , y).
\eea
The factorization formula for the jet-charge-binned 1-Jettiness distribution is now given by
\bea
\label{eq:fac_resum_Q_J_Rq_bin}
&&d\sigma_{\rm resum} \left [{\cal Q}^{\rm bin}, \tau_1,P_{J_T},y_J \right ] =\sigma_0 \> H(\xi, \mu; \mu_H)  \\
&&\times \int ds_J\>\int dt_a \> \int ds_J'  \>{\cal S}\left(\tau_1 - \frac{t_a}{Q_a}-\frac{s_J}{Q_J}, \mu;\mu_S\right) 
\nn \\
&&\times \Big [ \sum_{q_i} \>L_{q_i}\>R_{\rm bin}^{q_i}(s_J-s_J')J^{q_i}(s_J', \mu;\mu_J)  B_{q_i}(x_*,t_a,\mu;\mu_B)  \nn \\
&&\>\>\>+ \sum_{\bar{q}_i} L_{\bar{q}_i}\>R^{\bar{q}_i}_{\rm bin}(s_J-s_J')J^{\bar{q}_i}(s_J', \mu;\mu_J)  \> B_{\bar{q}_i}(x_*,t_a,\mu;\mu_B) \Big ] . \nn
\eea

\section{1-Jettiness Jet Charge Distribution with Polarized Protons}
\label{polprotons}

The 1-Jettiness Jet Charge formalism can be straightforwardly extended to DIS with longitudinally polarized protons. The 1-Jettiness distribution of the longitudinally polarized nucleon asymmetry serves as a probe of the helicity PDFs. The ${\cal Q}$-binned 1-Jettiness distribution then allows enhanced flavor separation of the helicity PDFs. 

The resummation formula for the 1-Jettiness distribution, $\tau_1$, of the longitudinally polarized nucleon asymmetry is schematically given by~\cite{Boughezal:2017tdd}
\bea
\label{schem-pol}
d\Delta \sigma_{\rm resum} \left [\tau_1, P_{J_T}, y_J \right ] \sim \Delta H\otimes \Delta B \otimes J \otimes {\cal S},
\eea
where $\Delta$ denotes the polarization dependence in hard and beam functions needed to obtain the longitudinal nucleon polarization asymmetry. The polarized beam function~\cite{Stewart:2010tn,Stewart:2009yx,Boughezal:2017tdd}, $\Delta B$, can be factorized as 
\bea
\label{schem-pol-2}
\Delta B\sim \Delta {\cal I}\otimes \Delta f,
\eea
where $\Delta {\cal I}$ is perturbatively calculable and $\Delta f$ denotes the helicity PDF. The corresponding factorization and resummation formula for the 1-Jettiness jet charge distribution, in addition to the standard 1-Jettiness distribution, is given by replacing the standard jet function, $J$, with the charged jet function, ${\cal G}$:
\bea
\label{schem-pol}
d\Delta \sigma_{\rm resum} \left [{\cal Q}, \tau_1, P_{J_T}, y_J \right ] \sim \Delta H\otimes \Delta B \otimes {\cal G} \otimes {\cal S}.
\eea
Note that the charged jet function ${\cal G}$ that appears here is identical to the one that appears in the 1-Jettiness Jet Charge formalism with unpolarized protons, corresponding to its universality property. Similarly, studying the jet-charge-binned 1-Jettiness distribution of the longitudinally nucleon spin asymmetry facilitates disentangling the flavor structure of the helicity PDFs. We leave detailed phenomenological studies of the flavor structure of the helicity PDFs using the 1-Jettiness jet charge formalism for future work. In this work, we focus on introducing the main ideas and formalism of 1-Jettiness jet charge in DIS of unpolarized protons only.

\section{Universality of the Charged Jet Function}
\label{univ}

Eqs.~(\ref{eq:fac_resum_Q_J_Rq}) and (\ref{eq:fac_resum_Q_J_Rq_bin}) are key results, providing factorization formulae for the 1-Jettiness Jet Charge distribution. They give the fully differential cross section $d\sigma\left [Q_J, \tau_1,P_{J_T},y_J \right ]$ and the jet-charge-binned cross section  $d\sigma\left [Q_J^{\rm bin}, \tau_1,P_{J_T},y_J \right ]$, respectively, in the resummation region $\tau_1 \ll P_{J_T}$. The predictive power of this factorization framework is enhanced due to the universality of the various functions appearing in the factorization formulae. The hard function $H(\xi,\mu)$ describes universal QCD corrections to the hard scattering quark current vertex due to single photon or Z-boson exchange in standard fully inclusive DIS. It is perturbatively calculable with results known~\cite{Idilbi_2006,Becher:2006mr} to ${\cal O}(\alpha_s^2)$. The universal quark beam functions~\cite{Stewart:2009yx}, $B_q(x,t,\mu)$, describe the physics of perturbative initial state radiation, collinear with the proton beam direction, and the initial state collinear PDF, as seen in Eq.~(\ref{beam}). The quark beam functions are also known~\cite{Stewart:2010qs,Gaunt:2014xga} to ${\cal O}(\alpha_s^2)$. The jet function describes the physics of the energetic radiation in the jet region, and collinear with the leading jet direction, and is known~\cite{Becher:2006qw, Idilbi_2006,Becher:2006mr} to ${\cal O}(\alpha_s^2)$. Note that for the jet functions initiated by light quarks, the light quark flavor symmetry of QCD in the limit of massless quarks, leads to the additional universality, $J^{q_i}(s,\mu)=J^{q}(s,\mu)$, so that the jet functions are independent of quark flavor.
 The soft function ${\cal S}(\tau_1,\mu)$ describes the soft radiation, $E\sim \tau_1$, throughout the event. \red{In the resummation region, 
 $\tau_1\ll P_{J_T}$, the soft function is perturbatively calculable when $ \tau_1\gg \Lambda_{\rm QCD}$. When $ \tau_1\sim \Lambda_{\rm QCD}$  the soft function becomes non-perturbative and requires the introduction of a non-perturbative model. In the tail region, $ \tau_1\sim P_{J_T}\gg \Lambda_{\rm QCD}$, it can be expanded~\cite{Korchemsky:1998ev,Korchemsky:1999kt,Bauer:2003di,Lee:2006fn,Lee:2006nr} through an operator product expansion (OPE) in powers of $\Lambda_{\rm QCD}/\tau_1$. The first term in the OPE is perturbatively calculable and the dominant non-perturbative effects come from the first correction term in the OPE.  However, the soft function, and its associated  non-perturbative effects, are universal since it is the same soft function that appears in Eq.~(\ref{eq:factorization_resum}) for the standard 1-Jettiness distribution where no jet charge measurement is made. }

The new ingredients in the 1-Jettiness Jet Charge factorization formula are the $R^q(Q,s)$  functions appearing in Eq.~(\ref{eq:fac_resum_Q_J_Rq}).  In general, they are non-perturbative objects since the jet charge measurement is not collinear safe and sensitive to the hadronization process. However, these non-perturbative functions are universal and  renormalization scale independent. They can be extracted from other observables that also make jet charge measurements on the jet regions of global event shapes. N-Jettiness~\cite{Stewart:2010tn} is an ideal global event shape for such purposes. It characterizes how closely the  pattern of final state radiation resembles a N-jet like event. Small values of N-Jettiness, $\tau_N$, correspond to events that closely resemble an exclusive N-Jet event. Large values of $\tau_N$ correspond to events with additional jets or energetic radiation in different directions. The resummation region corresponds to small values of N-Jettiness, $\tau_N\ll Q$, where $Q$ denotes the hard scale in the problem. For $e^+e^-$ colliders, in the resummation region, the N-Jettiness differential distribution has the schematic form
\bea
\label{NJettiness-1}
\frac{d\sigma_{\rm resum}}{d\tau_N} \sim H\otimes  J \otimes \cdots \otimes J \otimes S,
\eea
where $H$ describes the hard scattering process, $J$ denotes the standard jet function, and there are N such factors each corresponding to one of the N jet directions,  and $S$ denotes the soft function that describes soft radiation throughout the event. Note that the case of N=2 can be related to the thrust global event in $e^+e^-$ colliders in the limit $\tau_2\ll Q$. 

In order to access the universal ${\cal G}^q(Q,s,\mu)$ jet charge function, and the corresponding $R^q(Q,s)$ functions needed for the 1-Jettiness Jet Charge distribution, one can generalize the N-Jettiness observables at $e^+e^-$ and hadron colliders to include a jet charge measurement on one or more of the jet regions. For example, if a jet charge measurement is made on the jet region with the highest energy $E_J$ or transverse momentum $P_{J_T}$, the formula in Eq.~(\ref{NJettiness-1})  becomes
\bea
\label{JCNJettiness-1}
\frac{d\sigma_{\rm resum}}{d\tau_Nd{\cal Q}} \sim H\otimes  {\cal G} \otimes \cdots \otimes J \otimes S .
\eea
The factor of the jet charge jet function, ${\cal G}$, replaces the standard jet function, $J$, in the jet region where the jet charge measurement is made. These formulae at $e^+e^-$  colliders can be compared to Eq.~(\ref{schem-3}) for the 1-Jettiness Jet Charge distribution, and the same universal jet charge jet function, ${\cal G}$, appears in all these observables. 

One can also consider the N-Jettiness distributions in a give jet charge bin, where we integrate over the jet charge of measured jet region. The corresponding jet-charge-binned N-Jettiness distributions for $e^+e^-$ colliders take the form
\bea
\label{JCNJettiness-1-Bin}
\frac{d\sigma_{\rm resum}^{{
\cal Q}\text{-}\rm bin} }{d\tau_N} \sim H\otimes  {\cal G}_{\rm bin} \otimes \cdots \otimes J \otimes S.
\eea
Here ${\cal G}_{\rm bin}$ denotes the jet charge jet function integrated over a jet charge bin and is the same as the one that appears in Eq.~(\ref{schem-3-Bin}) for the jet-charge-binned 1-Jettiness distribution in electron-proton collisions.

Furthermore, one can also consider different definitions of the N-Jettiness event shapes, associated with the choice of parameters in their definitions. For example, the dimensionless $\tau_{1a}$ event shape~\cite{Kang:2013nha,Kang:2014qba,Chu:2022jgs,Cao:2024ota} differs from the definition of $\tau_1$ in Eq.~(\ref{tau1}) in the choice of the $Q_{J,B}$ values. Similarly, one can work with dimension one $\tau_N$ or dimensionless $\mathcal{T}_N$. For all of these different definitions,  corresponding to different partitioning of final state radiation into beam and jet regions, the same universal jet functions or  charged jet functions appear in the factorization formulae.

Thus, the jet charge jet functions ${\cal G}^q$ and ${\cal G}^q_{\rm bin}$, and correspondingly the functions $R^q({\cal Q},s)$ and $R^q_{\rm bin}(s)$ are universal and can be extracted from a comprehensive and systematic global analysis of the wide range of N-Jettiness global event shape observables at $e^+e^-$  colliders.

\section{Non-Perturbative Model for Charged Jet  Function  }
In order to calculate the 1-Jettiness jet charge distribution, we need to quantify the  nonperturbative component of the charged jet function ${\cal G}^{q}({\cal Q},y,\mu)$, or equivalently, its Fourier transform, ${\cal G}^{q}({\cal Q},s,\mu)$. From Eqs.~(\ref{Rq}) and (\ref{JCFconv}), this corresponds to specifying the nonperturbative functions, $R^q({\cal Q},y)$, and their Fourier transform functions, $R^q({\cal Q},s)$.  In the next three subsections, we present the general parameterization of a phenomenological model for $R^q({\cal Q},y)$ and $R^q({\cal Q},s)$ and consider the simplest case of these function being independent of $y$ or $s$, as well as a phenomenological parameterization of additional $y$ or $s$ dependence beyond the simplest model.

\subsection{General Model Parameterization}

The form of the nonperturbative functions, $R^q({\cal Q},y)$ and $R^q({\cal Q},s)$, is constrained by the normalization conditions in Eq.~(\ref{Rqnorm}). Given any arbitrary function ${\cal K}^q({\cal Q},y)$, the normalization condition on $R^q({\cal Q},y)$ can be satisfied by constructing it from ${\cal K}^q({\cal Q},y)$ as follows,
\bea
\label{rqyform}
R^q({\cal Q},y) =\frac{{\cal K}^q({\cal Q},y)}{\int d{\cal Q}\>{\cal K}^q({\cal Q},y)},
\eea
as long as $\int d{\cal Q}\>{\cal K}^q({\cal Q},y)\neq 0$. The corresponding $R^q({\cal Q},s)$ function is obtained via the Fourier transform of Eq.~(\ref{rqyform}) with respect to the $y$ variable, as seen in Eq.~(\ref{RqFT})
\bea
\label{rqsform}
R^q({\cal Q},s) =\int \frac{dy}{2\pi} e^{iys} \left [\frac{{\cal K}^q({\cal Q},y)}{\int d{\cal Q}\>{\cal K}^q({\cal Q},y)} \right ],
\eea
from which it follows that $R^q({\cal Q},s)$ satisfies the required normalization condition in Eq.~(\ref{Rqnorm}).

\subsection{Simple Model Parameterization}

In this section, we consider the simplest case where the charged jet functions ${\cal G}^q({\cal Q},y,\mu)$ or ${\cal G}^q({\cal Q},s,\mu)$ \red{have} the same $y$ or $s$ dependence as the standard jet functions, $J^q(y,\mu)$ or $J^q(s,\mu)$, respectively. This corresponds to the $R^q({\cal Q},y)$ being independent of $y$, as seen from Eqs.~(\ref{RqRatio}). Equivalently, $R^q({\cal Q},s)$  will be proportional to $\delta(s)$, which follows from Eq.~(\ref{RqFT}). 
In this special case, the $R^q({\cal Q},y)$  and $R^q({\cal Q},s)$ functions can be written as
\bea
\label{Rqapprox}
R^{q}({\cal Q},y) &=& r^q({\cal Q}),\qquad R^{q}({\cal Q},s) = r^q({\cal Q}) \delta(s), 
\eea
where the function $r^q({\cal Q})$ satisfies the normalization condition
\bea
\label{rq_chi_norm}
\int d{\cal Q} \> r^q({\cal Q}) =1.
\eea
From Eqs.~(\ref{Rq}) and (\ref{JCFconv}), the charged jet functions then reduce to the separable form
\bea
\label{Gqapprox}
{\cal G}^{q}({\cal Q},y,\mu) &=& r^q({\cal Q})\>J^{q}(y,\mu), \nn \\
{\cal G}^{q}({\cal Q},s,\mu) &=& r^q({\cal Q})\>J^{q}(s,\mu),
\eea
where the charged jet function, ${\cal G}$, and the ordinary jet function, $J$, having the same dependence in the $y$ or $s$ variables. 
Using this model in Eqs.~(\ref{Rqapprox}) and (\ref{Gqapprox}),  the factorization formula in Eq.~(\ref{eq:fac_resum_Q_J_Rq}) can be written as
\bea
\label{eq:fac_resum_Q_J_Rq_Approx}
&&d\sigma_{\rm resum} \left [{\cal Q}, \tau_1,P_{J_T},y_J \right ] =\sigma_0 \> H(\xi, \mu; \mu_H) \nn \\
&&\times \int ds_J\>\int dt_a  \>{\cal S}\left(\tau_1 - \frac{t_a}{Q_a}-\frac{s_J}{Q_J}, \mu;\mu_S\right) 
 \\
&&\times \Big [ \sum_{q_i} \>L_{q_i}\>r^{q_i}({\cal Q})J^{q_i}(s_J, \mu;\mu_J)\>  B_{q_i}(x_*,t_a,\mu;\mu_B)  \nn \\
&&+ \sum_{\bar{q}_i} L_{\bar{q}_i}\>r^{\bar{q}_i}({\cal Q}) J^{\bar{q}_i}(s_J, \mu;\mu_J)\> B_{\bar{q}_i}(x_*,t_a,\mu;\mu_B) \Big ] , \nn
\eea
where the jet charge information is now contained entirely in the $r^q({\cal Q})$ functions.  The jet-charge-binned cross section in Eq.~(\ref{eq:fac_resum_Q_J_Rq_bin}) then takes the following form
\bea
\label{eq:fac_resum_Q_J_Rq_Approx_Bin}
&&d\sigma_{\rm resum} \left [{\cal Q}_J^{\rm bin}, \tau_1,P_{J_T},y_J \right ] =\sigma_0 \> H(\xi, \mu; \mu_H) \nn \\
&&\times \int ds_J\>\int dt_a  \>{\cal S}\left(\tau_1 - \frac{t_a}{Q_a}-\frac{s_J}{Q_J}, \mu;\mu_S\right) 
 \\
&&\times \Big [ \sum_{q_i} \>L_{q_i}\>r^{q_i}_{\rm bin}J^{q_i}(s_J, \mu;\mu_J)\>  B_{q_i}(x_*,t_a,\mu;\mu_B)  \nn \\
&&+ \sum_{\bar{q}_i} L_{\bar{q}_i}\>r^{\bar{q}_i}_{\rm bin} J^{\bar{q}_i}(s_J, \mu;\mu_J)\> B_{\bar{q}_i}(x_*,t_a,\mu;\mu_B) \Big ] , \nn
\eea
where the $r^{q}_{\rm bin}$  constants are given by the integration of the $r^q({\cal Q})$ function over the jet charge bin
\bea
\label{rqbinint}
r^q_{\rm bin} = \int_{\rm bin} d{\cal Q} \> r^q({\cal Q}),
\eea
and satisfy the constraint
\bea
\label{rqbinunity}
\sum_{{\rm bins}} r^q_{\rm bin} = 1,
\eea
for each quark and anti-quark flavor, $q$, as required by Eq.~(\ref{rq_chi_norm}). Thus, the $r^q_{\rm bin}$ values can be interpreted as the fraction of jets initiated by a parton of flavor $q$ in the given jet charge bin. Based on the PYTHIA 8.312 simulations of five million events of the unpolarized DIS process, $e^-+p\to e^- + J  + X$, for the typical EIC kinematics of $\sqrt{s}=90.0$ GeV, $P_{J_T}= [20.0,30.0]$ GeV, and $y_J=[-2.5,2.5]$, we show in Tables~\ref{tab:StdJCBinFrac} and \ref{tab:DynJCBinFrac} the  $r^q_{\rm bin}$ bin fractions for the Standard and Dynamic jet charge definitions, respectively. The $r^q_{+,-,0}$ bin fractions correspond to the fraction of jets in the positive $({\cal Q} > 0.25)$, negative $( {\cal Q} < -0.25)$, and central $(-0.25 < {\cal Q} < 0.25)$ jet charge bins, respectively. 
\begin{table}[]
    \centering
    \begin{tabular}{|c|cccccc|}
    \hline
     & $u$ & $\bar{u}$ &  $d$ & $\bar{d}$ & $s$ & $\bar{s}$ \\
    \hline
       $r^i_+$  & 0.62 & 0.18 &0.13&0.50&0.19&0.52 \\
        $r^i_-$ & 0.08 & 0.50 &0.48&0.13&0.43& 0.13 \\
        $r^i_0$ &0.30 & 0.32 &0.39&0.37&0.38&0.35  \\
        \hline
    \end{tabular}
    \caption{Standard Jet Charge binned fractions for different quark flavors.  The relevant EIC kinematics chosen are: $\sqrt{s}=90.0$ GeV, $P_{J_T}= [20.0,30.0]$ GeV, and $y_J=[-2.5,2.5]$.}
    \label{tab:StdJCBinFrac}
\end{table}

\begin{table}[]
    \centering
    \begin{tabular}{|c|cccccc|}
    \hline
     & $u$ & $\bar{u}$ &  $d$ & $\bar{d}$ & $s$ & $\bar{s}$ \\
    \hline
       $r^i_+$  & 0.39 & 0.15 &0.15&0.29&0.15&0.35 \\
        $r^i_-$ & 0.11 & 0.36 &0.29&0.15&0.33& 0.14 \\
        $r^i_0$ &0.49 & 0.49 &0.55&0.55&0.52&0.51  \\
        \hline
    \end{tabular}
    \caption{Dynamic Jet Charge binned fractions for different quark flavors.  The relevant EIC kinematics chosen are: $\sqrt{s}=90.0$ GeV, $P_{J_T}= [20.0,30.0]$ GeV, and $y_J=[-2.5,2.5]$.}
    \label{tab:DynJCBinFrac}
\end{table}

We can write the jet-charge-binned 1-Jettiness Jet Charge cross section in the resummation region as
\bea
\label{rqbin_cross}
&&d\sigma_{\rm resum}\big [{\cal Q}^{\rm bin}, \tau_1, P_{J_T},y_J \big ]  \\
 &&=\sum_q r^q_{\rm bin}\>d\sigma_{\rm resum}^q \left [ \tau_1,P_{J_T},y_J \right ], \nn
\eea
where the sum over $q$ runs over all the quark and anti-quark flavors. The cross section $d\sigma_{\rm resum}^q \left [\tau_1,P_{J_T},y_J \right ]$ corresponds to the contribution from the $q$-parton flavor to the standard 1-Jettiness distribution:

\bea
\label{eq:fac_resum_q}
&&d\sigma_{\rm resum}^q \left [\tau_1,P_{J_T},y_J \right ] =\sigma_0 \> L_{q}\> H(\xi, \mu; \mu_H) \nn \\
&&\times \int ds_J \int dt_a \>J^q(s_J, \mu;\mu_J)  B_{q}(x_*,t_a,\mu;\mu_B) 
\nn \\
&&\times   \> \> {\cal S}\left(\tau_1 - \frac{t_a}{Q_a}-\frac{s_J}{Q_J}, \mu;\mu_S\right) ,
\eea
while $r^q_{\rm bin}$ determines the weight given to this contribution based on the fraction of jets initiated by the corresponding parton flavor in the given jet charge bin.

\begin{figure}
    \includegraphics[scale=0.5]{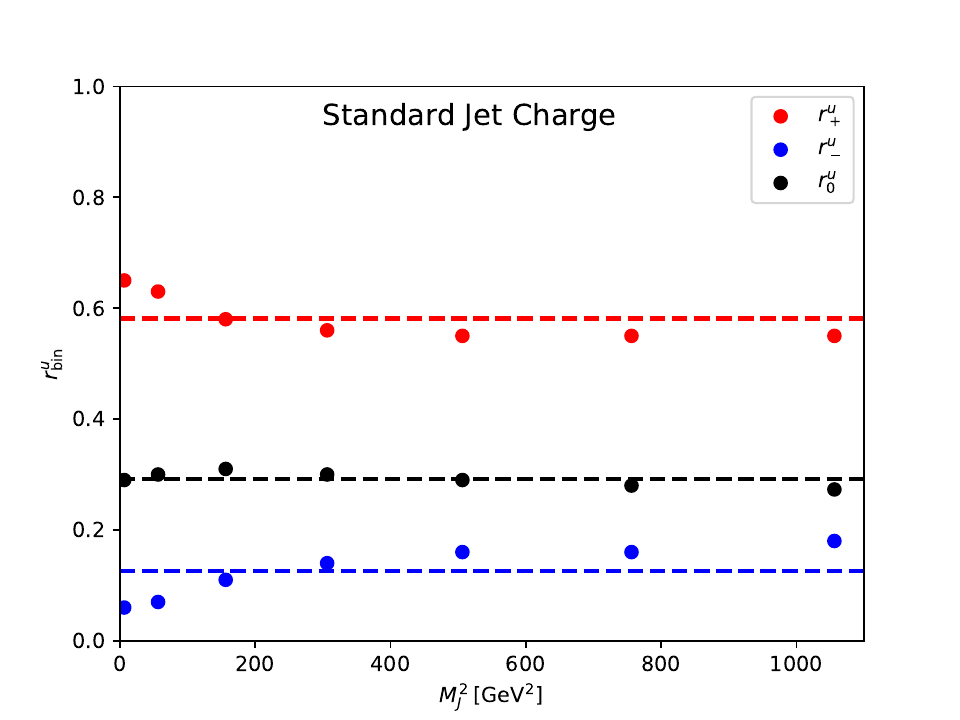}
        \includegraphics[scale=0.5]{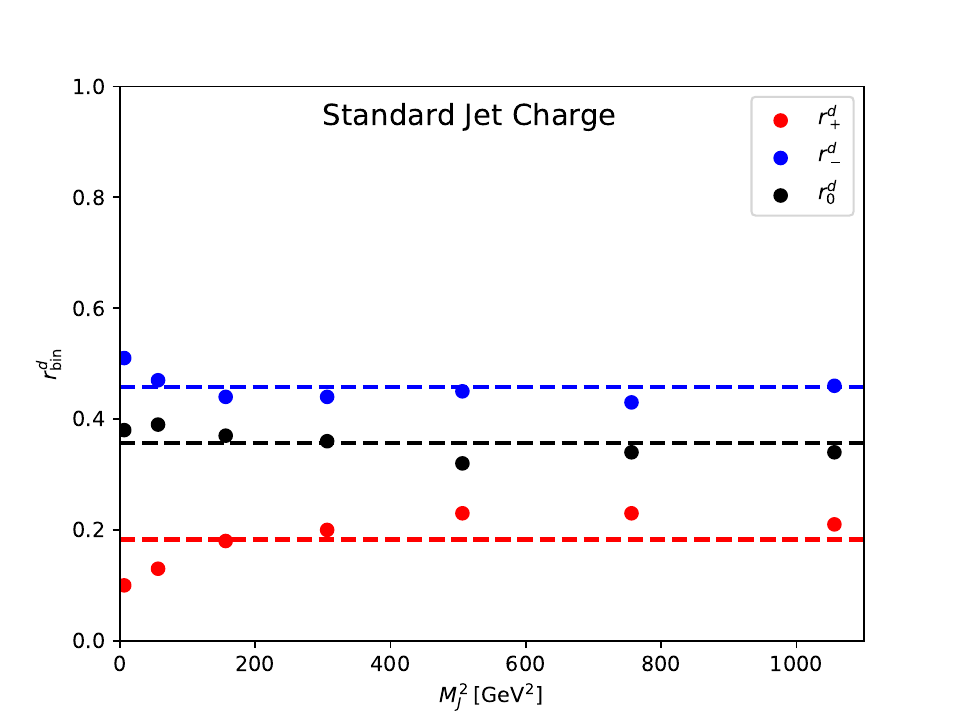}
    \caption{The fraction of jets initiated by the $u$-quark  (top panel) and $d$-quark (bottom panel) in the positive $(Q_J>0.25)$, negative $(Q_J<-0.25)$, and central $(-0.25 < Q_J<0.25)$ Standard Jet Charge bins as a function of the squared jet mass, $M_J^2$.  The dashed lines indicate the corresponding average values from the fraction of jets over all squared jet mass bins. The relevant EIC kinematics chosen for generating the Pythia 8.312 simulation data are: $\sqrt{s}=90.0$ GeV, $P_{J_T}= [20.0,30.0]$ GeV, and $y_J=[-2.5,2.5]$. }
    \label{fig:rqMJstd}
\end{figure}

\begin{figure}
    \includegraphics[scale=0.5]{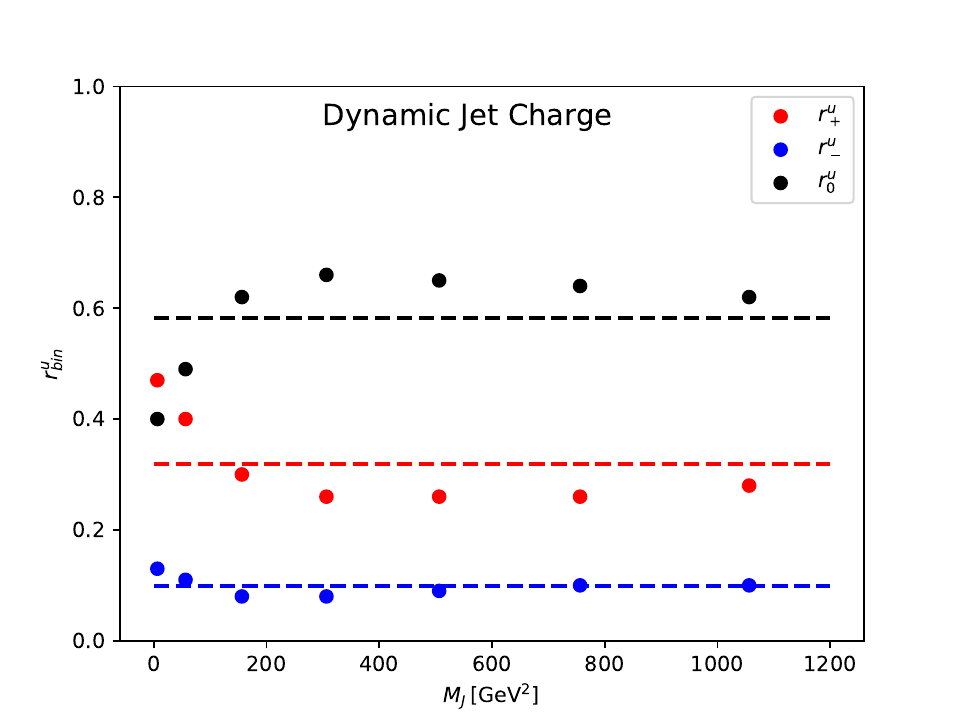}
        \includegraphics[scale=0.5]{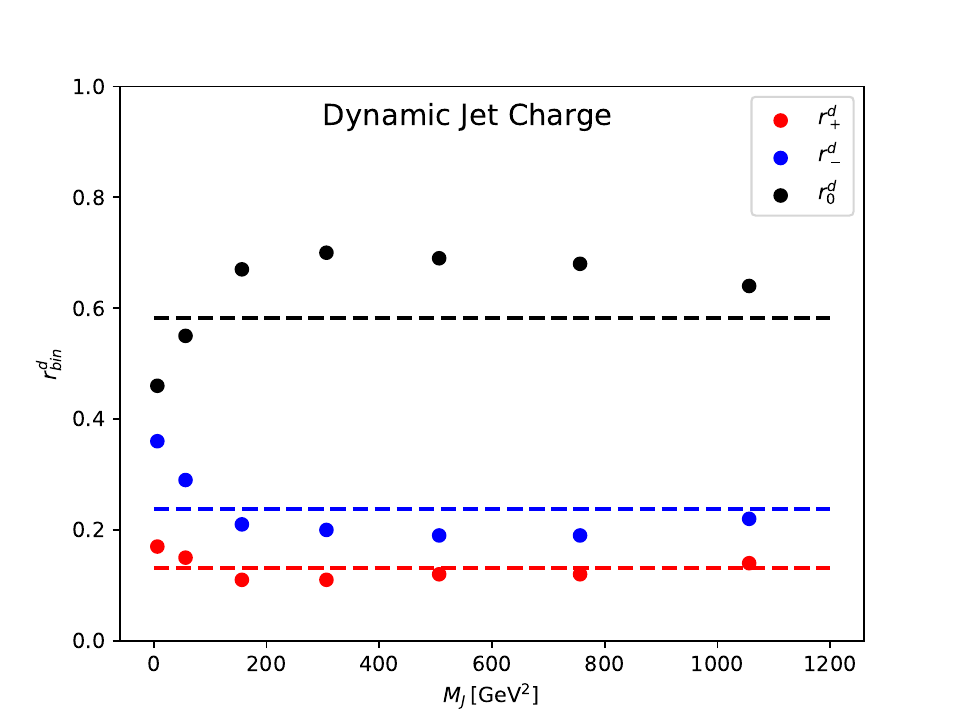}
    \caption{The fraction of jets initiated by the $u$-quark  (top panel) and $d$-quark (bottom panel) in the positive $(Q_J>0.25)$, negative $(Q_J<-0.25)$, and central $(-0.25 < Q_J<0.25)$ Dynamic Jet Charge bins as a function of the squared jet mass, $M_J^2$.  The dashed lines indicate the corresponding average values from the fraction of jets over all squared jet mass bins. The relevant EIC kinematics chosen for generating the Pythia 8.312 simulation data are: $\sqrt{s}=90.0$ GeV, $P_{J_T}= [20.0,30.0]$ GeV, and $y_J=[-2.5,2.5]$. }
    \label{fig:rqMJdyn}
\end{figure}

We note that the form of Eq.~(\ref{rqbin_cross}) is similar to that in Ref.~\cite{Kang:2020fka} which studied $e^-+p\to e^-+J+X$, in the limit of small electron-jet transverse momentum imbalance, $q_T=|\vec{p}_T^{\>e}+\vec{P}_{J_T}|\ll\vec{p}_T^{\>e} \sim P_{J_T}$,  in different jet charge bins to probe the flavor structure of TMDPDFs. There the jet was defined with a jet algorithm and fixed jet radius. In our case, where the jet charge binning is applied to a jet region of a global event shape, the form in Eq.~(\ref{rqbin_cross}) only applies in the special limit of Eq.~(\ref{Gqapprox}) where the charged jet function has the same $y$ or $s$ dependence as the standard jet function. The more general parameterization allows for a more non-trivial dependence of the charged jet function, ${\cal G}(s,\mu)$ and ${\cal G}(y,\mu)$ on the variables $s$ or $y$ that goes beyond the dependence in the standard jet function $J(s,\mu)$ and $J(y,\mu)$, respectively. In general, this leads to the non-trivial convolution structure in Eqs.~(\ref{eq:fac_resum_Q_J_Rq}) or (\ref{eq:fac_resum_Q_J_Rq_bin}).

\subsection{A Useful Model Parameterization}

The form of the jet-charge-binned 1-Jettiness cross section in Eq.~(\ref{rqbin_cross}) is the direct result of the special case of the nonperturbative model, as shown in Eq.~(\ref{Rqapprox}). In particular, the $r^q_{\rm bin}$ parameters are constants, independent of $s$. We test how well this model works using PYTHIA simulations. 
Note that the variable $s$ is the mass squared of all the energetic collinear particles in the jet region, arising from the charged jet function, ${\cal G}(s,\mu)$. However, the true measurable jet mass squared of the jet region also includes the contribution of soft particles in the jet region. The jet mass squared can be written as $M_J^2 = (p_c+p_s)^2$ where $p_c^\mu$ and $p_s^\mu$ denote the total sum of the momenta of the collinear and soft particles, respectively.  We can write $M_J^2 = s+2p_c\cdot p_s+p_s^2$, where $s=p_c^2$. We note that while the soft particles in the jet region make a significant contribution to jet mass, $M_J$, via the $2p_c\cdot p_s\sim s$ term, their contribution to the jet charge is suppressed since the contribution of each particle is weighed by its energy as seen in Eq.~(\ref{jc}). Thus, the fraction, $r^q_{\rm bin}$, of jets initiated by parton flavor, $q$, in a given jet charge bin is primarily determined by the energetic collinear particles, with the contributions from soft particles being power suppressed. Thus, the behavior of $r^q_{\rm bin}$ as a function of $M_J^2 = (p_c+p_s)^2$ will be similar to that as a function of $s=p_c^2$ such that we can write
\bea
r^q_{\rm bin}(M_J^2) \simeq r^q_{\rm bin}(s).
\eea
In Figs.~\ref{fig:rqMJstd} and \ref{fig:rqMJdyn}, we show the jet mass dependence of the jet charge bin fractions based on PYTHIA 8.312 simulations for $r^u_{+,-,0}(M_J^2) \simeq r^u_{+,-,0} (s)$ and $r^d_{+,-,0}(M_J^2) \simeq r^d_{+,-,0} (s)$  for the Standard and Dynamic jet charge definitions, respectively.  We again choose typical EIC kinematics with $\sqrt{s}=90.0$ GeV, $P_{J_T}= [20.0,30.0]$ GeV, and $y_J=[-2.5,2.5]$. 

 The results in Figs.~\ref{fig:rqMJstd} and \ref{fig:rqMJdyn} suggest that the $r^q_{\rm bin}$ do have a non-trivial dependence on $M_J^2$, and correspondingly suggest a non-trivial dependence on $s$. This dependence is stronger for the Dynamic Jet Charge compared to the Standard Jet Charge. These results shows that the simplified charged jet function model in Eq.~(\ref{Gqapprox}) which predicts constant $r^q_{\rm bin}$ values that are independent of $s$, is only an approximation and the more general model parameterization in Eqs.~(\ref{rqyform}) and (\ref{rqsform}) is appropriate, where ${\cal K}({\cal Q},y)$ has a non-trivial $y$-dependence.

 However, as seen in Figs.~\ref{fig:rqMJstd} and \ref{fig:rqMJdyn}, while the $r^q_{\rm bin}$ values are not constant, the dependence on $M_J^2$, or equivalently on $s$, is still mild. This suggests a model parameterization for $R^q({\cal Q},y)$  as a sum of a constant term and a $y$-dependent correction term as
 \bea
\label{Rqmody2}
R^{q}({\cal Q},y) &=& r^q({\cal Q}) + f^q({\cal Q},y). 
\eea
The $y$-dependent function $f^q({\cal Q},y)$ characterizes the deviation of $R^{q}({\cal Q},y)$ from the being a constant, independent of $y$ as characterized by the $r^q({\cal Q})$ term. Equivalently, as seen from Eq.~(\ref{Rq}), $f^q({\cal Q},y)$ describes the additional $y$-dependence in the charged jet function, ${\cal G}^q({\cal Q},y,\mu)$, compared to that of the standard jet function, $J^q(y,\mu)$. In order to satisfy the normalization condition on $R^{q}({\cal Q},y)$ in Eq.~(\ref{Rqnorm}), and given the normalization condition already satisfied by $r^q({\cal Q})$ in Eq.~(\ref{rq_chi_norm}), the function $f^q({\cal Q},y)$ must satisfy
\bea
\label{fqcond}
\int d{\cal Q} \> f^q({\cal Q},y) = 0.
\eea
Since this condition on $f^q({\cal Q},y)$ must be satisfied for all values of $y$, it must either be a function that is separable in the variables $y$ and ${\cal Q}$, or more generally, be built out of a sum of separable functions as
\bea
\label{fqbasis}
f^q({\cal Q},y) =  \sum_n g_n^q(y)\> \chi_n^q({\cal Q}).
\eea
The functions $\chi_n^q({\cal Q})$ can be chosen to form a complete set of  basis functions, satisfy the conditions
\bea
\label{rq_chi_norm_formal}
\int d{\cal Q} \>\chi_n^q({\cal Q}) &=&0 , \nn \\
\int d{\cal Q} \>\chi_n^q({\cal Q}) \> \chi_m^q({\cal Q}) &=& \delta_{m,n}.
\eea
The first condition ensures the required constraint in Eq.~(\ref{fqcond}). The second orthonormality condition allows the coefficient functions, $g^q(y)$, for a given $f^q({\cal Q},y)$, to be determined as
\bea
g_n^q(y) &=& \int d{\cal Q}\> f^q({\cal Q},y) \>\chi_n^q({\cal Q}). 
\eea 
 
In momentum space, the $R^q({\cal Q},s)$ function is parameterized as
 \bea
\label{Rqsmod_formal}
R^q({\cal Q},s) &=&  r^q({\cal Q})\> \delta(s) + f^q({\cal Q},s),
\eea
where  $f^q({\cal Q},s)$ are the Fourier transforms of $f^q({\cal Q},y)$
\bea
f^q({\cal Q},s) &=& \int \frac{dy}{2\pi} \> e^{i y s}\> f^q({\cal Q},y), 
\eea
and $f^q({\cal Q},s)$ satisfy the condition
\bea
\label{fqcond_s}
\int d{\cal Q}\>f^q({\cal Q},s) =0,
\eea
which follows from Eq.~(\ref{fqcond}), and is required by the underlying normalization condition in Eq.~(\ref{Rqnorm}). The momentum space function $f^q({\cal Q},s)$ can also be expressed as a sum over separable functions in terms of a complete set of basis functions as
\bea
\label{fqbasis_s}
f^q({\cal Q},s) =  \sum_n g_n^q(s)\> \chi_n^q({\cal Q}),
\eea
where the coefficient functions $g_n^q(s)$ are related to the $g_n^q(y)$ functions via the Fourier transform as 
\bea
g_n^q(s) &=& \int \frac{dy}{2\pi} \> e^{iys} g_n^q(y).
\eea
Having identified some of the key structural features of the nonperturbative functions, $R({\cal Q},y)$ and $R({\cal Q},s)$, we leave a more detailed model building analysis for future work. 
 
\section{Numerical Results}

In this section, we demonstrate the sensitivity of the 1-Jettiness Jet Charge, ${\cal Q}$, to probe quark flavor dynamics in nucleon structure \red{and} explore ways of probing the hadronization process. All the numerical results presented here are based on the PYTHIA 8.312 simulations of five million events of the unpolarized DIS process, $e^-+p\to e^- + J  + X$, for the typical EIC kinematics of $\sqrt{s}=90.0$ GeV, $P_{J_T}= [20.0,30.0]$ GeV, and $y_J=[-2.5,2.5]$. In this analysis, we only present results for unpolarized DIS which probe the unpolarized PDFs of the initial state nucleon. These results can be straightforwardly generalized to studies with longitudinally polarized initial state nucleons to probe the polarized PDFs, using the formalism described in section~\ref{polprotons}. 
\begin{figure}
    \centering
        \includegraphics[scale=0.5]{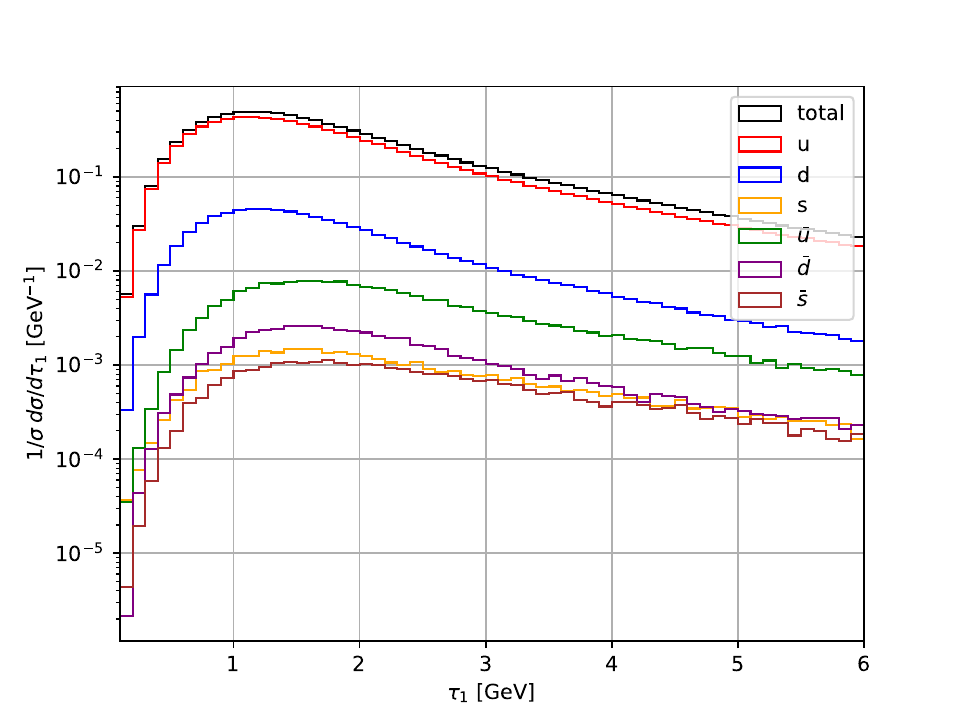}
    \caption{The normalized 1-Jettiness $\tau_1$-distribution (black) using Pythia data along with the relative contributions of different quark flavors. The relevant EIC kinematics chosen are: $\sqrt{s}=90.0$ GeV, $P_{J_T}= [20.0,30.0]$ GeV, and $y_J=[-2.5,2.5]$.}
    \label{Norm-One-Jettiness}
\end{figure}

\begin{figure}
    \centering
    \includegraphics[scale=0.5]{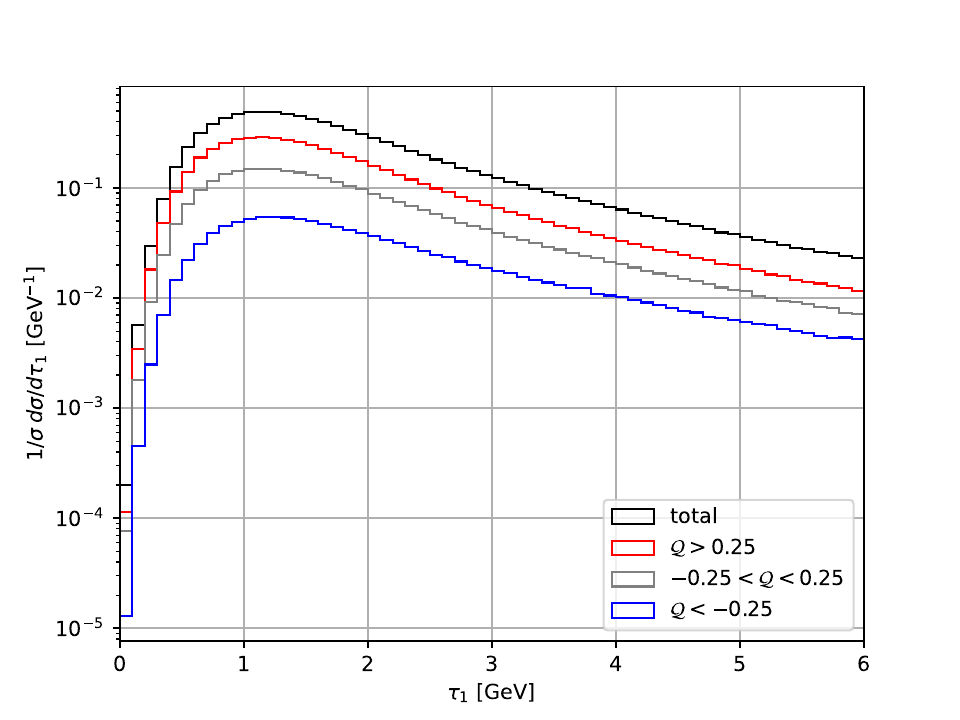}
    \caption{The normalized 1-Jettiness $\tau_1$-distribution (black) using Pythia data along with the relative contributions from the negative (blue), central (gray), and positive (red) standard jet charge bins corresponding to ${\cal Q} <-0.25$, $-0.25< {\cal Q} <0.25$, and ${\cal Q} >0.25$, respectuvely.
 The relevant EIC kinematics chosen are: $\sqrt{s}=90.0$ GeV, $P_{J_T}= [20.0,30.0]$ GeV, and $y_J=[-2.5,2.5]$. }
    \label{Norm-1-Jettiness-Binned-by-Std-Jet-Charge}
\end{figure}

\begin{figure}
    \centering
    \includegraphics[scale=0.45]{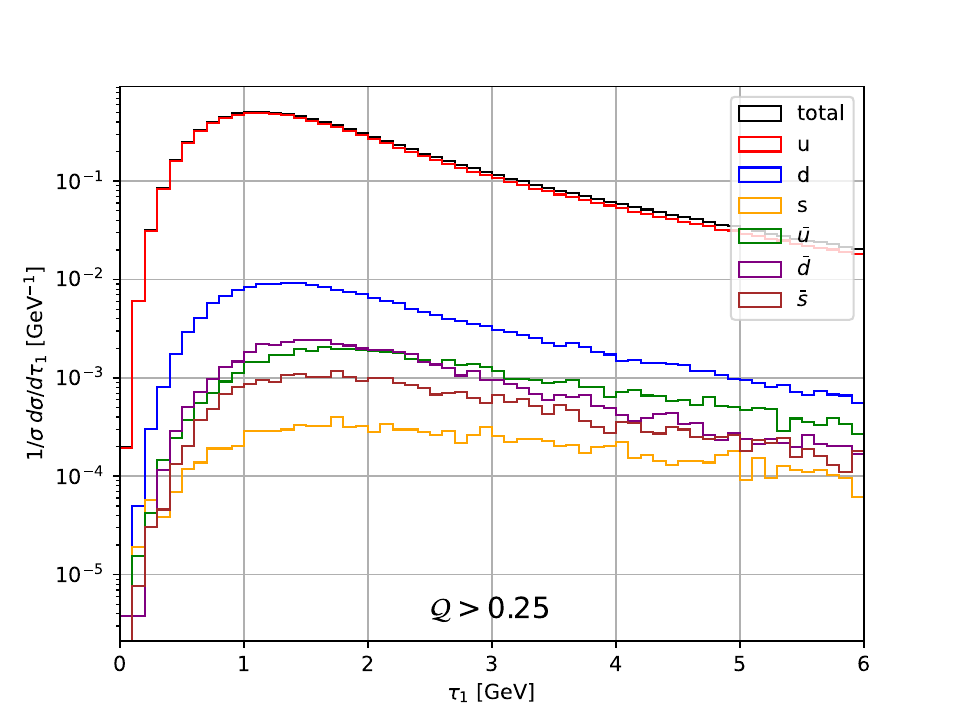}
       \includegraphics[scale=0.45]{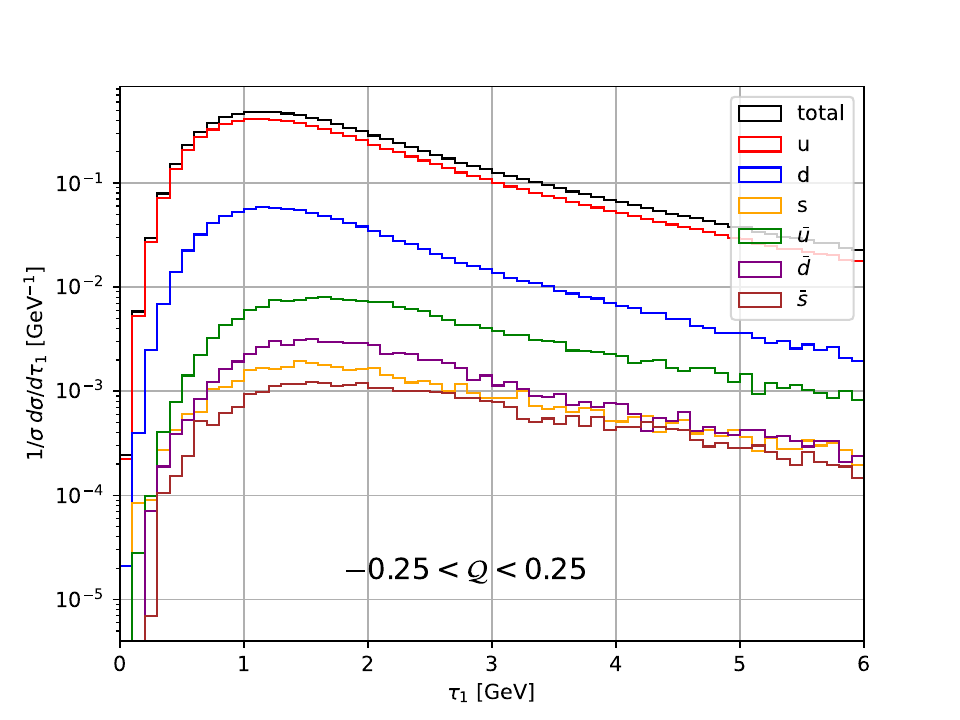}
    \includegraphics[scale=0.45]{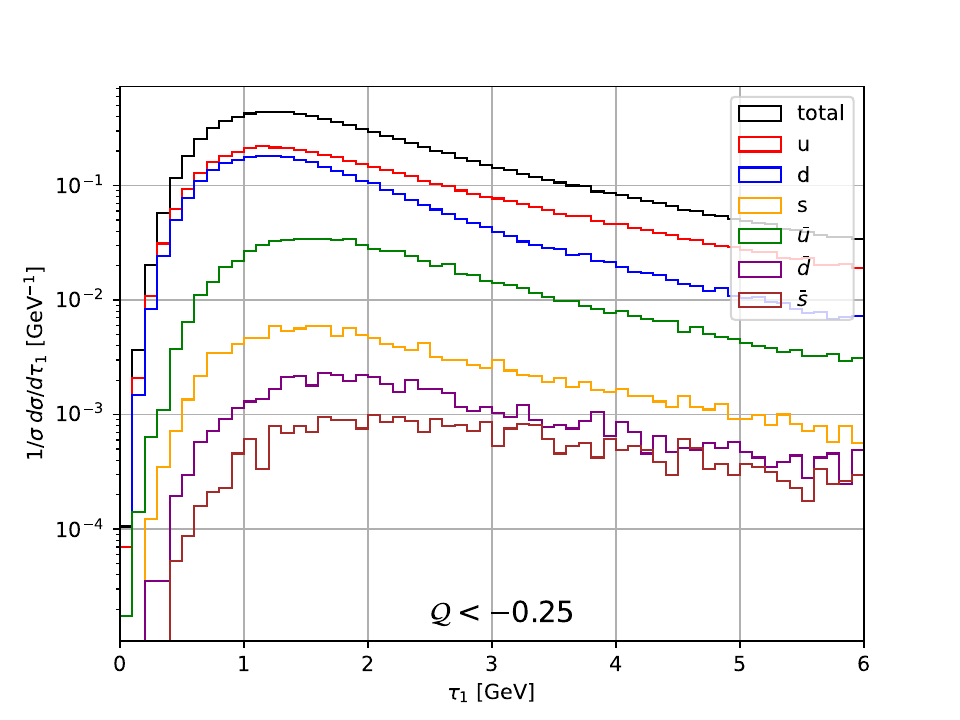}
    \caption{The normalized 1-Jettiness $\tau_1$-distribution (black) using Pythia data in the positive (${\cal Q}>0.25,$ top panel), central ($-0.25 < {\cal Q}<0.25,$ middle panel), and negative (${\cal Q}<-0.25,$ bottom panel) Standard Jet Charge bins, along  with the relative contributions of different quark flavors. 
 The relevant EIC kinematics chosen are: $\sqrt{s}=90.0$ GeV, $P_{J_T}= [20.0,30.0]$ GeV, and $y_J=[-2.5,2.5]$.}
    \label{stdJCbinJettiness}
\end{figure}

\begin{figure}
    \centering
    \includegraphics[scale=0.45]{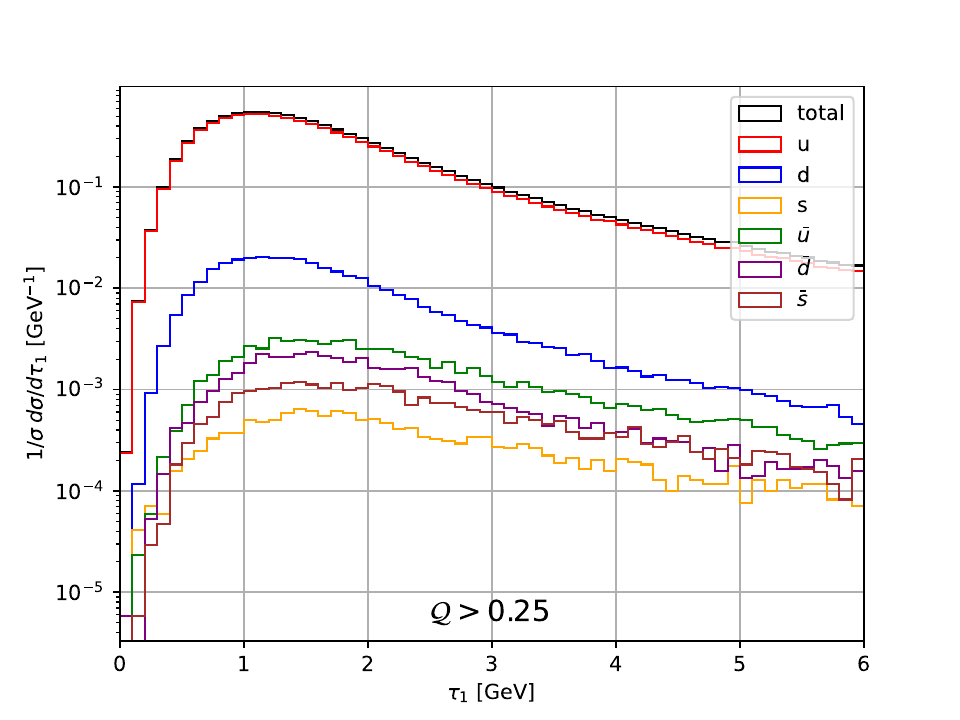}
        \includegraphics[scale=0.45]{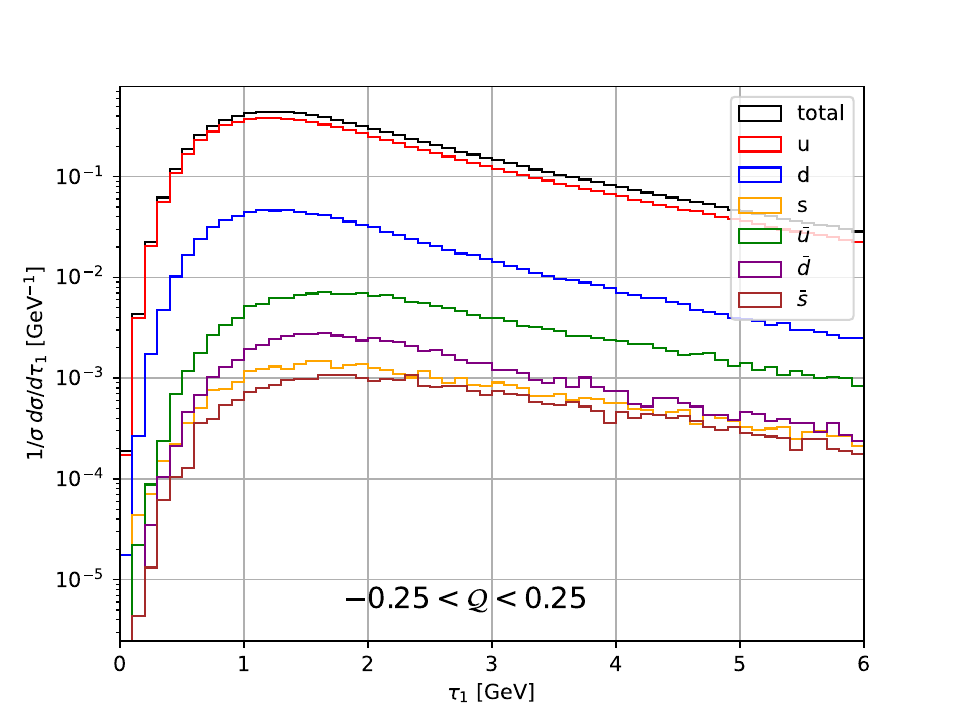}
    \includegraphics[scale=0.45]{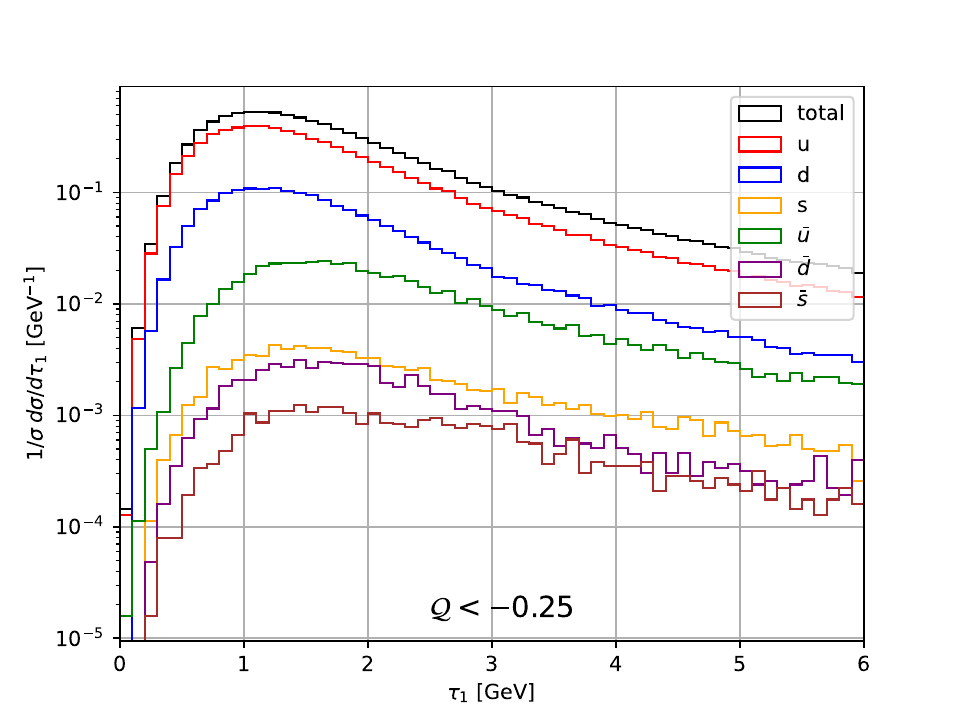}
    \caption{The normalized 1-Jettiness $\tau_1$-distribution (black) using Pythia data in the positive (${\cal Q}>0.25,$ top panel), the central ($-0.25 <{\cal Q}<0.25,$ middle panel) and negative (${\cal Q} <-0.25,$ bottom panel) Dynamic Jet Charge bins, along  with the relative contributions of different quark flavors. 
 The relevant EIC kinematics chosen are: $\sqrt{s}=90.0$ GeV, $P_{J_T}= [20.0,30.0]$ GeV, and $y_J=[-2.5,2.5]$.}
    \label{dynJCbinJettiness}
\end{figure}
We consider the standard jet charge definition with constant $\kappa =0.3$ in Eq.~(\ref{jc}). We will also present results based on the dynamic jet charge~\cite{Kang:2021ryr} definition, corresponding to $\kappa = \kappa (z_h,\xi_{\rm cut}, k_<, k_>)$ where
\begin{equation}
  \kappa (z_h) =\begin{cases}  
  k_<, & z_h < \xi_{\rm cut} \\
  k_> , & z_h \geq \xi_{\rm cut} \\
  \end{cases}
\label{eq:dynfuncform}  
\end{equation}
and $\xi_{\rm cut},~ k_<$, and $k_>$ are three constant parameters. We choose the default values of $\xi_{\rm cut}=0.3, \>k_< =1.0, \>k_>=0.3$. The standard and dynamic jet charge definitions have different jet discrimination capabilities and sensitivity to power corrections from soft radiation effects, thereby serving as complementary probes of quark flavor dynamics in nuclear structure and hadronization.

\subsection{Probing Nucleon Flavor Structure} 

In Fig.~\ref{Norm-One-Jettiness}, we show the normalized 1-Jettiness distribution (upper black curve). We also show the contributions to the 1-Jettiness distribution from scattering off quarks of different flavors in the initial state nucleon. We see that the 1-Jettiness distribution is overwhelmingly dominated by the $u$-quark flavor due to the dominance of its unpolarized PDF in the proton in the valence region. As a result, this distribution is not very sensitive to the dynamics of the $d$-quarks and the sea quarks and antiquarks. 

One can gain enhanced sensitivity to the $u$-antiquark and the $d$- and $s$-quark unpolarized PDFs by binning the 1-Jettiness distribution according the value of the  1-Jettiness Standard Jet Charge, ${\cal Q}$. In Fig.~\ref{Norm-1-Jettiness-Binned-by-Std-Jet-Charge}, we show the 1-Jettiness distribution binned according to ${\cal Q}$. We see the relative contribution to the total 1-Jettiness distribution from each jet charge bin. The dominant contribution comes from the positive $({\cal Q} > 0.25)$ jet charge bin, corresponding to the fact that the 1-Jettiness distribution is dominated by the contribution from the $u$-quark which initiates jets with a standard jet charge distribution peaked in the positive jet charge bin. Note that $u$-quark initiated jets still have a non-zero probability of having a standard jet charge value in the negative $({\cal Q} < - 0.25)$ or central $(-0.25 < {\cal Q} <  0.25)$ jet charge bins, as indicated by the fact that the contribution of the positive jet charge bin is not as dominant compared to the contribution from the $u$-quark in Fig.~\ref{Norm-One-Jettiness}.

In Fig.~\ref{stdJCbinJettiness}, we show the relative contributions to the 1-Jettiness distrubution from different quark flavors in each standard jet charge bin. We see that in the positive (top panel) and central (middle panel) jet charge bin, the 1-Jettiness distribution is overwhelmingly dominated by the $u$-quark contribution. However, we see that in the negative (bottom panel) jet charge bin, the contribution from the $d$-quark is comparable to that from the $u$-quark, indicating that one has enhanced sensitivity to the $d$-quark PDF when the 1-Jettiness distribution is restricted to the negative jet charge bin. In Fig.~\ref{dynJCbinJettiness}, we show the corresponding results using the dynamic jet charge definition. Once again, we see enhanced sensitivity to the $d$-quark PDF in the negative dynamic jet charge bin. Similarly, studying the 1-Jettiness distribution for DIS on longitudinally polarized protons, according to  jet charge bins can help disentangle the flavor structure of the helicity PDFs.  In this manner, the 1-Jettiness jet charge measurement can be used as a new probe of the quark flavor structure in the nucleon. 

\subsection{Probing Flavor Dynamics in Hadronization}

\begin{figure}
    \includegraphics[scale=0.5]{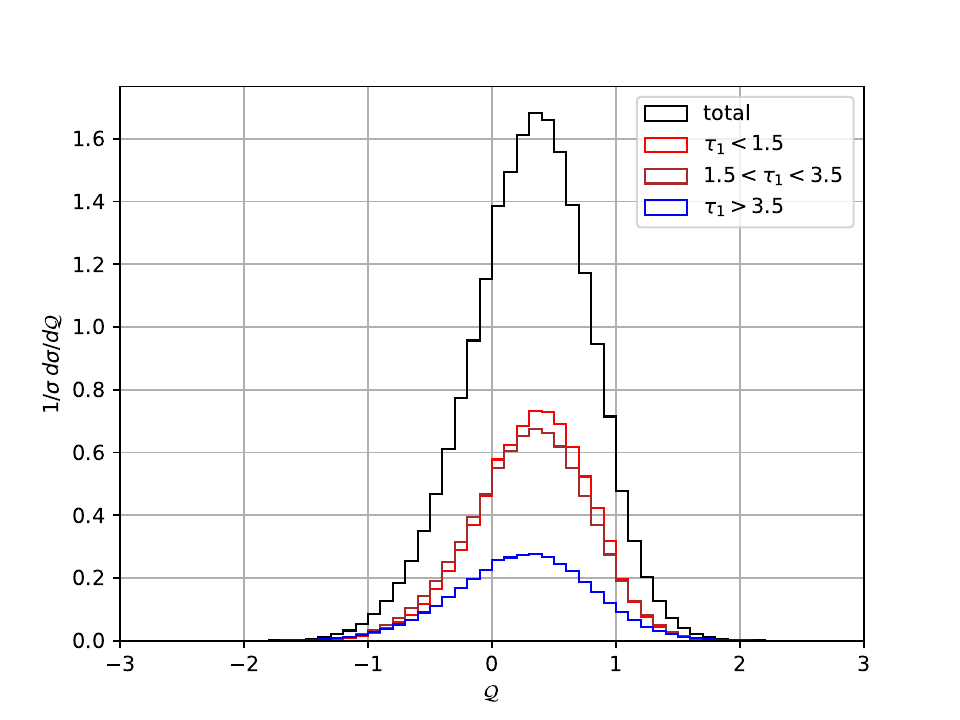}
        \includegraphics[scale=0.5]{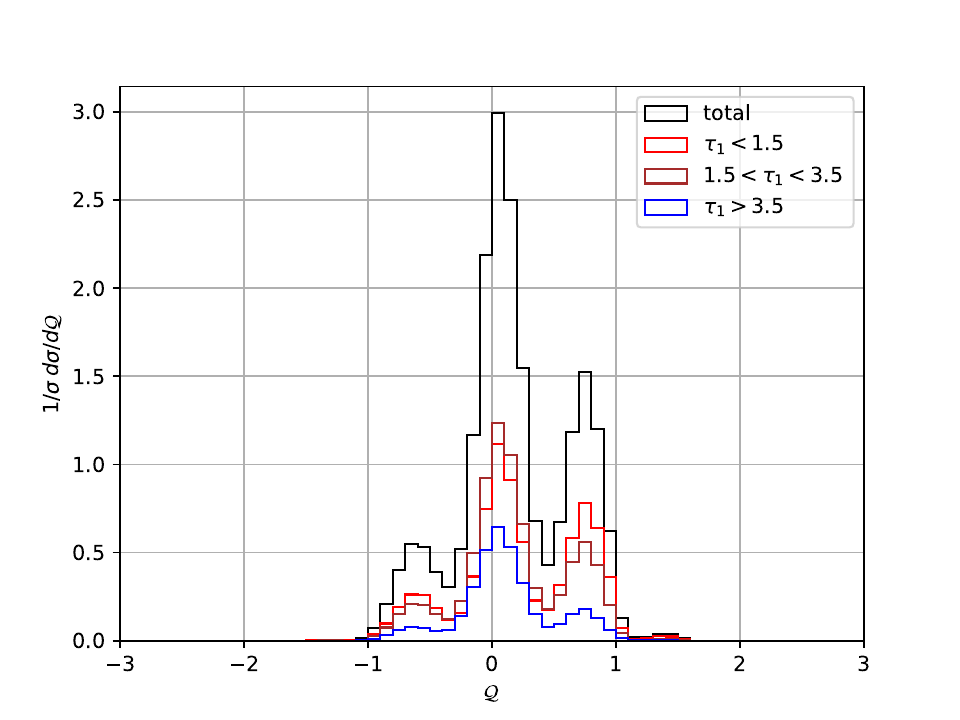}
    \caption{1-Jettiness Jet Charge distributions according to 1-Jettiness bins for the Standard Jet Charge (top panel) and the Dynamic Jet Charge (bottom panel). The binned curves are normalized to the total unbinned jet charge distribution curve which in turn is normalized to unity over the displayed range.  The relevant EIC kinematics chosen are: $\sqrt{s}=90.0$ GeV, $P_{J_T}= [20.0,30.0]$ GeV, and $y_J=[-2.5,2.5]$.}
    \label{QJdisttau1bin}
\end{figure}

\begin{figure}
    \includegraphics[scale=0.5]{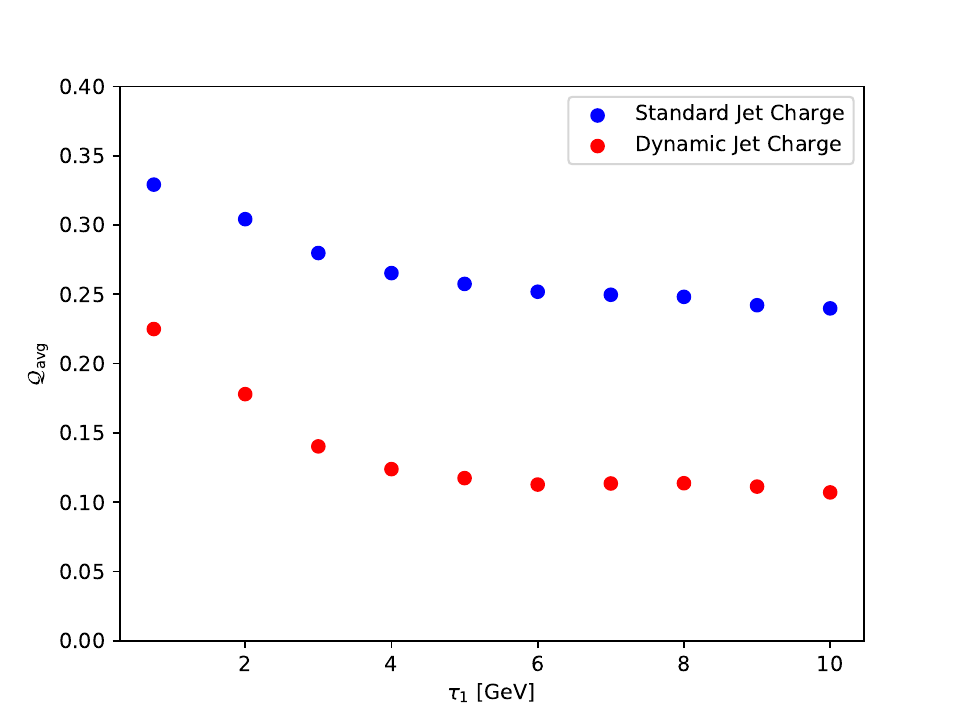}
        \includegraphics[scale=0.5]{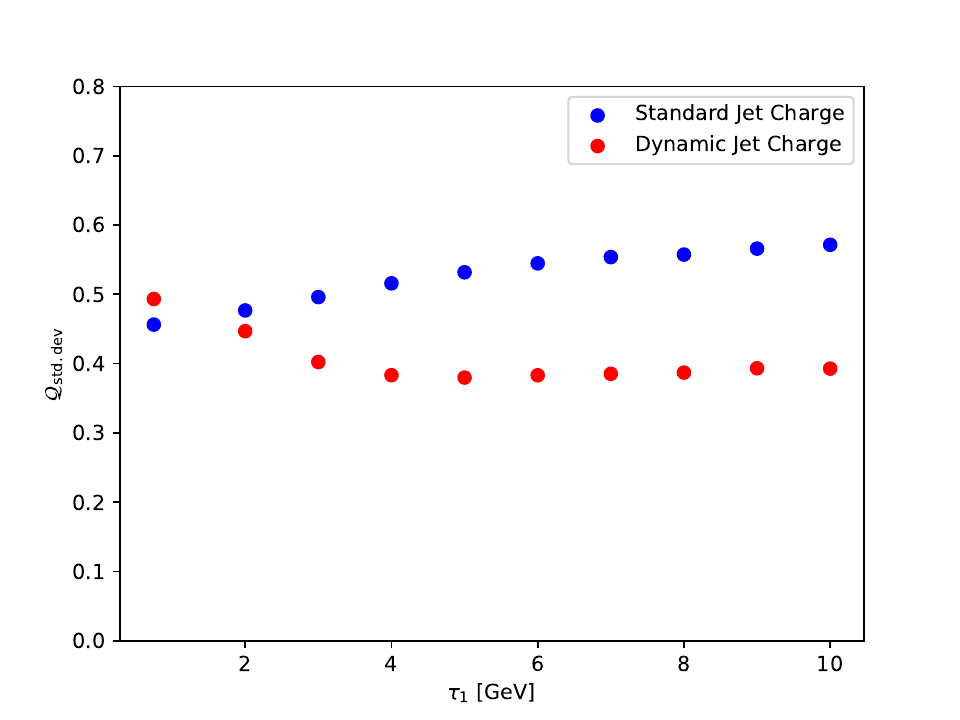}
    \caption{The average (top panel) and standard deviation (bottom panel) values of jet charge distributions binned according to the value of 1-Jettiness, $\tau_1$.  The relevant EIC kinematics chosen are: $\sqrt{s}=90.0$ GeV, $P_{J_T}= [20.0,30.0]$ GeV, and $y_J=[-2.5,2.5]$.}
    \label{fig:QJtau1avgstddev}
\end{figure}

One can also consider the 1-Jettiness jet charge distribution binned according to the 1-Jettiness value characterizing the shape of the energy flow in each event. This allows us to study the jet charge, ${\cal Q}$, as a function of the shape of the overall energy flow as characterized by $\tau_1$.  Such studies can provide constraints on hadronization models, allowing us to probe correlations between flavor dynamics and energy flow. We use the same PYTHIA8.312 simulation data for the EIC kinematics of $\sqrt{s}=90.0$ GeV, $P_{J_T}= [20.0,30.0]$ GeV, and $y_J=[-2.5,2.5]$ to explore observables that can probe hadronization models. 

In Fig.~\ref{QJdisttau1bin}, we show the total ${\cal Q}$-distribution (black) and the ${\cal Q}$-distribution binned (colored) in different $\tau_1$-bins for the standard (top panel) and dynamic (bottom panel) jet charge definitions. The total unbinned ${\cal Q}$-distribution (black) is normalized to unity. The $\tau_1$-binned ${\cal Q}$-distributions in turn are normalized to the unbinned distribution so that we can see the relative fraction of events in the different $\tau_1$-bins, characterizing the pattern of energy flow, as a function of the jet region charge. 

Another potentially interesting observable is the average jet charge as a function of 1-Jettiness. Theoretically, it can be defined via the probability distribution in the 1-Jettiness jet charge, ${\cal Q}$:
\bea
P({\cal Q}, \tau_1^{\rm bin}, P_{J_T}^{\rm bin}, y_J^{\rm bin})  = \frac{d\sigma[{\cal Q}, \tau_1^{\rm bin}, P_{J_T}^{\rm bin}, y_J^{\rm bin}]}{d\sigma[\tau_1^{\rm bin}, P_{J_T}^{\rm bin}, y_J^{\rm bin}]},
\eea
where the ${\rm bin}$ superscript indicates an integration over the corresponding bin. Note that \red{$d\sigma[{\cal Q}, \tau_1, P_{J_T}, y_J]$} and \red{$d\sigma[\tau_1, P_{J_T}, y_J]$} are defined in Eqs.~(\ref{eq:obs0QJ}) and (\ref{eq:obs0}), respectively. This probability distribution satisfies the required normalization condition
\bea
\int d{\cal Q}\> P({\cal Q}, \tau_1^{\rm bin}, P_{J_T}^{\rm bin}, y_J^{\rm bin})  =1.
\eea
This probability distribution can be used to define moments of the jet charge. The $n$-th moment is defined as
\bea
\langle {\cal Q}^n \rangle = \int d{\cal Q}\> {\cal Q}^n \>P({\cal Q}, \tau_1^{\rm bin}, P_{J_T}^{\rm bin}, y_J^{\rm bin}), 
\eea
with the first moment corresponding to the average jet charge
\bea
{\cal Q}_{\rm avg.}(\tau_1^{\rm bin}, P_{J_T}^{\rm bin}, y_J^{\rm bin}) = \langle {\cal Q} \rangle .
\eea 
Similarly, the standard deviation of the jet charge is in any bin is given by
\bea
{\cal Q}_{\rm std. dev}(\tau_1^{\rm bin}, P_{J_T}^{\rm bin}, y_J^{\rm bin}) =\sqrt{\langle {\cal Q}^2 \rangle - \langle {\cal Q} \rangle^2  }. 
\eea
Using PYTHIA 8.312 simulation data, in Fig.~\ref{fig:QJtau1avgstddev} we show the  jet region charge average (top panel) and standard deviation (bottom panel) as a function of $\tau_1$ for the standard (blue) and dynamic (red) jet charge definitions. We see that the standard and dynamic jet charge definitions give rise to different behavior of the jet charge moments as a function of $\tau_1$. Such correlations between the pattern of energy flow, characterized by the $\tau_1$ global event shape, and the jet region  charge, ${\cal Q}$, could probe hadronization models. In particular, the distribution of various moments of the 1-Jettiness jet charge with different jet charge definitions, and over a wide range of $(\tau_1, P_{J_T}, y_J)$ bins could provide non-trivial constraints on hadronization models. We leave a more detailed study exploring these distributions for future work.

\section{Conclusions}

We have proposed bringing together the flavor discrimination power of the jet charge observable with global event shapes that characterize the shape and pattern of energy flow throughout the event. This framework provides a new way to probe the quark flavor structure of the nucleon and the quark flavor dynamics in the hadronization process. It allows for a measurement of the jet region charge while simultaneously keeping track of the energy flow throughout the event, as characterized by the global event shape.

We focused on the measurement of the 1-Jettiness jet charge (${\cal Q}$),  the jet charge of the jet region ($J$) defined within the  framework of the 1-Jettiness global event shape ($\tau_1$) for the Deep Inelastic Scattering (DIS) process, $e^- + p \to e^- + J + X$, with unpolarized or longitudinally polarized protons.  The 1-Jettiness distribution, $\tau_1$, binned according to jet charge, ${\cal Q}$, allows for  enhanced quark flavor separation of the initial state unpolarized or polarized PDFs. Alternatively, the jet charge distribution binned by 1-Jettiness serves as a probe of quark flavor dynamics in the final state hadronization process. We derived a factorization theorem that simultaneously measures  $\tau_1$ and ${\cal Q}$ in the resummation region, $\tau_1 \ll P_{J_T}$, where the correlation between the jet charge and the charge of the struck quark is the strongest ($P_{J_T}$ denotes the transverse momentum of the jet region). We also discussed the formulation of power corrections to this factorization framework. 
We considered both the standard and dynamic jet charge definitions in our analysis, serving as complementary probes with different jet discrimination capabilities and sensitivity to power corrections from soft radiation.

The factorization theorem gives rise to new universal and field theoretically well-defined \textit{charged jet functions} that describe the evolution of the struck quark into a final state of given  jet mass and charge values.  The charged jet functions are non-perturbative but have the same renormalization group evolution as the ordinary jet function.
We developed a general model parametrerization for the non-perturbative component of these charged jet functions and studied its general structure and properties. We described how these universal functions can be extracted from a wide range of N-Jettiness global event shape observables at $e^+e^-$ colliders, including the thrust global event shape. 
We provided numerical results through simulations studies using PYTHIA 8.312 to demonstrate the sensitivity of this observable to probe flavor dynamics in nucleon structure and explore the possibility of probing quark flavor dynamics in the final state hadronization process.  

The 1-Jettiness jet charge framework can be applied to longitudinal nucleon spin asymmetries to probe the flavor structure of helicity PDFs. It can also be applied to charged current DIS, providing additional handles on constraining quark flavor dynamics in the nucleon and the hadronization process.  This observable is ideally suited for the proposed EIC and can be applied to existing HERA data. We leave such further theoretical and phenomenological studies of the 1-Jettiness jet charge observable for future work.

\section{Acknowledgements}

The work of YTC is supported in part by the US Department of Energy (DOE) Contract No. DEAC05-06OR23177, under which Jefferson Science Associates, LLC operates Jefferson Lab,
and by the Department of Energy Early Career Award grant DE-SC0023304. SM thanks Jefferson Lab for their hospitality and support through the Visiting Faculty Program (VFP) during which part of this work was carried out.

\appendix

\section{Charged Jet Function}
\label{JCJetFunc}

\red{The field-theoretic definition of the standard jet function~\cite{Bauer:2003pi,Becher:2006qw, Idilbi_2006,Becher:2006mr,Bruser:2018rad}, $J^{q_i}(s,\mu)$, is given by
\bea
\label{Jqfield-theoretic}
&& \delta^{ab}\left [\frac{\nslash}{2} 
\right ]_{\beta\alpha} J^{q_i}(s,\mu) =\frac{1}{\pi}\sum_{X} (2\pi)^4\delta(\omega-\bar{n}\cdot P_{X}) \delta^{2}(P_X^\perp) \nn \\
  &&\times \langle 0|\chi_{n,\beta}^{q_i,a} |X\rangle\langle X |\bar{\chi}_{n,\alpha}^{q_i,b} |0\rangle \delta (s-\omega\> n\cdot P_X) ,
\eea
where $\chi_{n}^{q_i,a}=(W^\dagger_n \xi_n^{q_i})^a$  is the collinear jet field after decoupling from the soft fields. The index $a$ corresponds to the color index in the fundamental representation. We are using the standard  SCET~\cite{Bauer:2000ew,Bauer:2000yr,Bauer:2001ct,Bauer:2001yt,Bauer:2002nz,Beneke:2002ph} notation. For brevity in notation, we have set $n_J^\mu=n^\mu$ in the above equations, dropping the $J$ subscript. Thus, the collinear jet field $\chi_{n}^{q_i}$ and the perpendicular direction of $P_X^\perp$, should be understood to be defined with respect to the jet direction. }

\red{The charged jet function, ${\cal G}^{q_i}({\cal Q},s,\mu)$, is defined with an additional insertion of a jet charge measurement function $\delta ({\cal Q}-{\cal Q}_X)$ in the above expression
\bea
\label{Gqfield-theoretic}
&&\delta^{ab}\left [\frac{\nslash}{2} 
\right ]_{\beta\alpha}  {\cal G}^{q_i}({\cal Q},s,\mu) = \frac{1}{\pi}\sum_{X} (2\pi)^4\delta(\omega-\bar{n}\cdot P_{X}) \delta^{2}(P_X^\perp) \nn \\
&&\times \langle 0| \chi_{n,\beta}^{q_i,a} |X\rangle\langle X |\bar{\chi}_{n,\alpha}^{q_i,b} |0\rangle \delta (s-\omega\> n\cdot P_X)\delta ({\cal Q}-{\cal Q}_X) ,
\eea
where $Q_X$ denotes the total jet charge of the state $|X\rangle$ 
\bea
\hat{{\cal Q}} | X\rangle = Q_X |X\rangle , \qquad \langle X | \hat{{\cal Q}}^\dagger =  \langle X | Q_X ,
\eea
and the jet charge operator is Hermitian, $\hat{{\cal Q}}^\dagger=\hat{{\cal Q}}$, with real eigenvalues.}

\section{Power Corrections to 1-Jettiness Jet Charge}
\label{PC}

The factorization formula in Eq.~(\ref{eq:factorization_resum_Q_J}), in the resummation region $\tau_1\ll P_{J_T}$, will receive power corrections from the contribution of soft particles to the jet charge. Since the jet charge definition includes a weight of the energy fractions of the particles in the jet region, as seen in Eq.~(\ref{jc}), the contribution of soft particles ($z_h\sim \tau_1/P_{J_T}$) to the jet charge is formally suppressed compared to that of the collinear particles ($z_h\sim 1$)  by $(\tau_1/P_{J_T})^\kappa$. Thus, for given choices of the $\tau_1$ region and the parameter, $\kappa$,  the effects of the power corrections can be suppressed by choosing a large enough value of $P_{J_T}$. We also note that the effects of power corrections are expected to be further suppressed in the dynamic jet charge definition where $\kappa =\kappa(z_h)$ is a dynamic function of $z_h$ that allows one further suppress the contributions of soft particles, compared to the standard jet charge definition. In fact, it has been shown~\cite{Kang:2021ryr}  that the dynamic jet charge is very insensitive to soft radiation by noting that the jet discrimination power of the dynamic jet charge is essentially the same in pp-collisions and PbPb-collisions even though the latter have significantly more soft contamination from MPI. By contrast, some degradation in the discrimination power was observed for the standard jet charge in PbPb-collisions compared to pp-collisions. 

In this appendix, we articulate the general form of the power correction due to the contribution to the jet region charge from soft particles, in the resummation region, $\tau_1\ll P_{J_T}$.  The 1-Jettiness jet charge, ${\cal Q}$, measurement is incorporated by the insertion the delta function $\delta({\cal Q} - {\cal Q}_c^J -{\cal Q}_s^J)$ in the 1-Jettiness factorization framework that requires that the sum of the jet charge contributions from the collinear and soft particles add up to the measured jet region charge, ${\cal Q}$. Here ${\cal Q}_c^J$ and ${\cal Q}_s^J$ denote the contribution from the collinear and soft particles in the jet region. Correspondingly, the delta functions $\delta({\cal Q}_c^J-\hat{{\cal Q}}^{J}_c)$ and $\delta({\cal Q}_s^J-\hat{{\cal Q}}^{J}_s)$ are inserted in the field-theoretic matrix element definitions of the jet function and soft function to obtain the charged jet function, ${\cal G}^{q}({\cal Q}_c^J, s, \mu)$, and charged soft function ${\cal S}\left(\tau_1, {\cal Q}_s^J,\mu \right)$, respectively.  Here $\hat{{\cal Q}}^{J}_c$ and $\hat{{\cal Q}}^{J}_s$ are the jet region charge operators that act on the collinear and soft particles in the jet region to obtain the charges ${\cal Q}_c^J$ and ${\cal Q}_s^J$, respectively. For the charged jet function, this is seen in Eq.~(\ref{Gqfield-theoretic}). 

The standard soft function is given in terms of the generalized hemisphere soft function~\cite{Kang:2012zr, Kang:2013wca, Kang:2013nha} as
\bea
\label{soft-1}
&&{\cal S}\left(\tau_1, \mu \right) = \int dk_a \int dk_J \>\delta (\tau_1-k_a-k_J)\nn \\
 &&\qquad \qquad\qquad \qquad \qquad \times {\cal S} (k_a,k_J,\mu),
 \eea
where hemisphere soft function is given by
\bea
\label{QFT-definition}
{\cal S}(k_a,k_J,\mu) &=& \frac{1}{N_c}\sum_{X_s} \>\text{Tr}\>\langle 0 | \bar{T} [Y_{n_A}^\dagger Y_{n_J} ] (0) \nn \\
&&\times \delta(k_a - \frac{q_A \cdot K_{X_s}^{(a)}}{Q_a})\delta(k_J - \frac{q_J\cdot K_{X_s}^{(J)}}{Q_J}) |X_s \rangle \nn \\
&\times& \langle X_s | T  [Y_{n_J}^\dagger Y_{n_A} ] (0) | 0 \rangle.
\eea
Here we have defined the total soft momentum in the jet and beam regions as
\bea
\label{psoft}
K_{X_s}^{(J)} &=& \sum_{k\in X_s} p_k \>\theta (\frac{2q_A\cdot p_k}{Q_a} - \frac{2q_J\cdot p_k}{Q_J} ), \nn \\
K_{X_s}^{(a)} &=& \sum_{k\in X_s} p_k \>\theta (\frac{2q_J\cdot p_k}{Q_J} - \frac{2q_A\cdot p_k}{Q_a} ),
\eea
respectively. The $Y_{n_J}$ and $Y_{n_A}$ denote soft Wilson lines that sum up soft eikonal emissions along the jet and beam directions, respectively.  Correspondingly, the charged soft function can be obtained via
\bea
\label{soft-1}
&&{\cal S}\left(\tau_1,{\cal Q}_s^J, \mu \right) = \int dk_a \int dk_J \>\delta (\tau_1-k_a-k_J)\nn \\
 &&\qquad \qquad\qquad \qquad \qquad \times {\cal S} (k_a,k_J,{\cal Q}_s^J, \mu),
\eea
where ${\cal S} (k_a,k_J,{\cal Q}_s^J, \mu)$ denotes the charged hemisphere function which involves the jet charge measurement operator $ \delta({\cal Q}_s^J - \hat{{\cal Q}}^{J}_s)$ acting on the soft particles in the jet  region as
\bea
\label{QFT-definition}
{\cal S}(k_a,k_J,\mu) &=& \frac{1}{N_c}\sum_{X_s} \>\text{Tr}\>\langle 0 | \bar{T} [Y_{n_A}^\dagger Y_{n_J} ] (0) \nn \\
&&\times \delta(k_a - \frac{q_A \cdot K_{X_s}^{(a)}}{Q_a})\delta(k_J - \frac{q_J\cdot K_{X_s}^{(J)}}{Q_J}) |X_s \rangle \nn \\
&\times& \langle X_s |  \delta({\cal Q}_s^J - \hat{{\cal Q}}^J_s)\>T  [Y_{n_J}^\dagger Y_{n_A} ] (0) | 0 \rangle.
\eea
We then also integrate over all possible values of ${\cal Q}_c^J$ and ${\cal Q}_s^J$, subject to the constraint $\delta({\cal Q} - {\cal Q}_c^J -{\cal Q}_s^J)$. This results in the general factorization formula
\bea
\label{eq:QJ_operator}
&&d\sigma_{\rm resum} \left [{\cal Q}, \tau_1,P_{J_T},y_J \right ] =\sigma_0 \> H(x_*  \sqrt{s} P_{J_T}e^{-y_J}, \mu; \mu_H) \nn \\
&&\int d{\cal Q}_c^J \int  d{\cal Q}_s^J \>\>\delta({\cal Q} - {\cal Q}_c^J -{\cal Q}_s^J)\nn \\
&&\times \int ds_J \int dt_a  \> {\cal S}\left(\tau_1 - \frac{t_a}{Q_a}-\frac{s_J}{Q_J}, {\cal Q}_s^J, \mu;\mu_S\right) 
\nn \\
&&\times \Big [ \sum_{q_i} \>L_{q_i}\>{\cal G}^{q_i}({\cal Q}_c^J, s_J, \mu;\mu_J)\>  B_{q_i}(x_*,t_a,\mu;\mu_B)  \nn \\
&&\>\>\>+ \sum_{\bar{q}_i} L_{\bar{q}_i}\>{\cal G}^{\bar{q}_i}({\cal Q}_c^J, s_J, \mu;\mu_J) \> B_{\bar{q}_i}(x_*,t_a,\mu;\mu_B) \Big ] .
\eea
Next, in the above factorization formula, we can expand $\delta({\cal Q} - {\cal Q}_c^J -{\cal Q}_s^J)$ in the limit $Q_s^J\ll Q_c^J$ as
\bea
\label{eq:Qexp}
&&\delta({\cal Q} - {\cal Q}_c^J -{\cal Q}_s^J)\nn \\ 
&=& \delta({\cal Q} - {\cal Q}_c^J) 
+ \delta'({\cal Q} - {\cal Q}_c^J) {\cal Q}_s^J +\cdots
\eea
where the ellipses denote terms suppressed by higher powers in ${\cal Q}_s^J$. The first term in Eq.~(\ref{eq:Qexp}), $\delta({\cal Q} - {\cal Q}_c^J) $, along with the normalization condition on the charged soft function
\bea
&&\int dQ_s^J \>{\cal S}\>\left(\tau_1 - \frac{t_a}{Q_a}-\frac{s_J}{Q_J}, {\cal Q}_s^J, \mu;\mu_S\right) \nn \\
&&={\cal S}\left(\tau_1 - \frac{t_a}{Q_a}-\frac{s_J}{Q_J}, \mu;\mu_S\right),
\eea
reproduces the leading power factorization formula in Eq.~(\ref{eq:factorization_resum_Q_J}). The first power correction comes from the second term in Eq.~(\ref{eq:Qexp}) and is given by
\bea
\label{eq:QJ_PC}
&&d\sigma_{\rm resum}^{\rm PC} \left [{\cal Q}, \tau_1,P_{J_T},y_J \right ] =\sigma_0 \> H(x_*  \sqrt{s} P_{J_T}e^{-y_J}, \mu; \mu_H) \nn \\
&&\times \int ds_J \int dt_a \int  d{\cal Q}_s^J \> Q_s^J\> {\cal S}\left(\tau_1 - \frac{t_a}{Q_a}-\frac{s_J}{Q_J}, {\cal Q}_s^J, \mu;\mu_S\right) \nn \\
&&\times \Big [ \sum_{q_i} L_{q_i}\int d{\cal Q}_c^J\> \delta'({\cal Q} - {\cal Q}_c^J)\>{\cal G}^{q_i}({\cal Q}_c^J, s_J, \mu;\mu_J)\nn \\
&& \qquad \times  B_{q_i}(x_*,t_a,\mu;\mu_B)  \\
&&\>\>\>+ \sum_{\bar{q}_i} L_{\bar{q}_i}\int d{\cal Q}_c^J\>\delta'({\cal Q} - {\cal Q}_c^J) \>{\cal G}^{\bar{q}_i}({\cal Q}_c^J, s_J, \mu;\mu_J) \nn \\
&&\qquad  \times B_{\bar{q}_i}(x_*,t_a,\mu;\mu_B) \Big ] . \nn
\eea
In this manner, power corrections from the contribution of soft particles in the jet region to the jet charge can be systematically incorporated. We note however that this power correction will also require incorporating a non-perturbative model for the new charged soft function. We also note that this power correction only involves the first jet charge moment of the new charged soft function, as seen in the second line of Eq.~(\ref{eq:QJ_PC}).

\bibliographystyle{h-physrev3.bst}
\bibliography{jetcharge}
\end{document}